\documentclass{article}
\usepackage{graphicx} 
\usepackage{todonotes}
\usepackage{tikz}
\usepackage{amsmath}
\usepackage{amsfonts}
\usepackage{longtable}
\usepackage{url}
\usepackage{amssymb}
\usepackage{hyperref}
\usepackage{booktabs}
\usepackage{lscape}
\usepackage{soul}
\usepackage{authblk} 
\usepackage[font=small]{caption}
\usepackage[labelformat=simple, font=small]{subcaption}

\usepackage{array}
\usepackage{comment}
\usepackage{rotating}
\usepackage{lmodern}
\usepackage[version=4]{mhchem} 

\usepackage[a4paper, total={6in, 8in}]{geometry}

\usepackage{float}
\usepackage{struktex}

\usepackage[section]{placeins}

\newcommand{\scale}{1}
\newcommand{\flowColor}{CarnationPink}

\newcommand{\pathOne}{CarnationPink}
\newcommand{\pathTwo}{Turquoise}
\newcommand{\pathThree}{ForestGreen}

\usepackage{mod}
\usetikzlibrary {arrows.meta}
\tikzset{
	abstractNetwork/.style={font=\scriptsize, scale=0.7},
	vertexBase/.style={circle, minimum size=1em},
	vertex/.style={vertexBase, circle, draw=black},
	hyperEdge/.style={rectangle, draw},
	arrow/.style={semithick, -{Stealth[round]}},
	inputMol/.style={draw=WildStrawberry, very thick},
	outputMol/.style={draw=ProcessBlue, very thick},
	borrowMol/.style={draw=ForestGreen, very thick},
	inputCofactor/.style={Gray},
	outputCofactor/.style={Gray}
}
\newtheorem{definition}{Definition}[section]
\usepackage[dvipsnames]{xcolor}
\usepackage{colortbl}
\usetikzlibrary{arrows.meta, positioning}
\newcommand{\mset}[1]{\left\{\!\!\left\{#1\right\}\!\!\right\}}
\usepackage{tikz}
\usetikzlibrary{shapes.geometric, calc}

\tikzstyle{startstop} = [rectangle, rounded corners, 
minimum width=3cm, 
minimum height=0.5cm,
text centered, 
draw=black]

\tikzstyle{io} = [trapezium, 
trapezium stretches=true, 
trapezium left angle=70, 
trapezium right angle=110, 
minimum width=3cm, 
minimum height=0.5cm, text centered, 
draw=black]

\tikzstyle{process} = [rectangle, 
minimum width=3cm, 
minimum height=0.5cm, 
text centered, 
text width=2.5cm, 
draw=black]

\tikzstyle{decision} = [diamond, 
minimum width=3cm, 
minimum height=0.5cm, 
text centered, 
draw=black]
\tikzstyle{arrow} = [thick,->,>=stealth]

\usepackage{mod}
\newcommand\preRuleContent{\scriptsize}

\newcommand{\powersetRuleScale}{0.5}
\newcommand{\powersetRuleRowSkip}{\par\medskip}

\newcommand{\powersetRuleBlock}[5]{%
    \begin{minipage}[t]{\textwidth}
        \centering
        \preRuleContent
        \dpoRule[scale=\powersetRuleScale]{#2}{#3}{#4}
        \par\smallskip{\small #1}
        #5
    \end{minipage}%
    \powersetRuleRowSkip
}

\usepackage{graphicx} 
\usepackage{todonotes}
\usepackage{tikz}
\usepackage{amsmath}
\usepackage{amsfonts}
\usepackage{longtable}
\usepackage{url}
\usepackage[dvipsnames]{xcolor}
\usepackage[T1]{fontenc}
\usepackage{lmodern}
\usetikzlibrary{arrows.meta, positioning}
\title{ Systematic pathway comparison on the powerset of rule-based biochemical systems} 
\author[1, 2, *]{Anne-Susann Abel}
\author[1, *]{Sissel Banke}
\author[1, 3, *]{Erika M. Herrera Machado}
\author[1]{Jakob Lykke Andersen}
\author[3]{Peter Dittrich}
\author[1]{Rolf Fagerberg}
\author[1, 4]{Daniel Merkle}
\date{\today}
\affil[1]{\small Department of Mathematics and Computer Science, University of Southern Denmark, Odense, Denmark}
\affil[2]{\small Institute for Theoretical Chemistry, University of Vienna, Vienna, Austria}
\affil[3]{\small Faculty of Mathematics and Computer Science, Friedrich Schiller University Jena, Jena, Germany}
\affil[4]{\small Algorithmic Cheminformatics Group, Faculty of Technology \& Center for Biotechnology (CeBiTec),
    Bielefeld University, Bielefeld, Germany}
\affil[*]{Equally contributing authors}

\begin{document}

\maketitle
\begin{abstract}

    Computational pathway design often focuses on evaluating selected pathways or optimizing fluxes in a fixed network, but gives less direct access to the combinatorial question of which other enzyme subsets of the network can support productive alternative pathways.
    A structured computational analysis of these networks can act as a valuable pre-step to the pathway design process.
    We present here a systematic approach for exploring biochemical pathway alternatives across enzyme subsets, using a computational methodology based on a rule-based modeling of the enzymes: for a biochemical system with enzyme set $S$, we evaluate subsets all $s \subseteq S$ by generating chemical reaction spaces, searching for integer-hyperflow pathways from prescribed inputs to target products, and organizing feasible subsets by set inclusion. This yields an inclusion-ordered landscape of pathway feasibility and carbon efficiency. We apply the approach to the non-oxidative pentose phosphate pathway, to non-oxidative glycolysis, and to glycolysis. Across these systems, feasible subsets occupy only a moderate fraction of all enzyme subsets, but the structure of this feasible region differs strongly between the systems. Larger enzyme sets do not consistently improve carbon efficiency when every enzyme in the tested subset is required to participate in the pathway. Instead, performance depends on specific enzyme combinations. The resulting subset landscapes are valuable means for identifying essential enzymes, candidate redundancies, and small high-performing enzyme subsets. By making the enzyme-subset landscape itself the object of analysis, the approach addresses the gap between detailed evaluation of individual candidate pathways and early-stage design decisions about which enzyme combinations are worth investigating at all.

    \begin{figure}[H]
    \centering
    \includegraphics[width=0.85\textwidth]{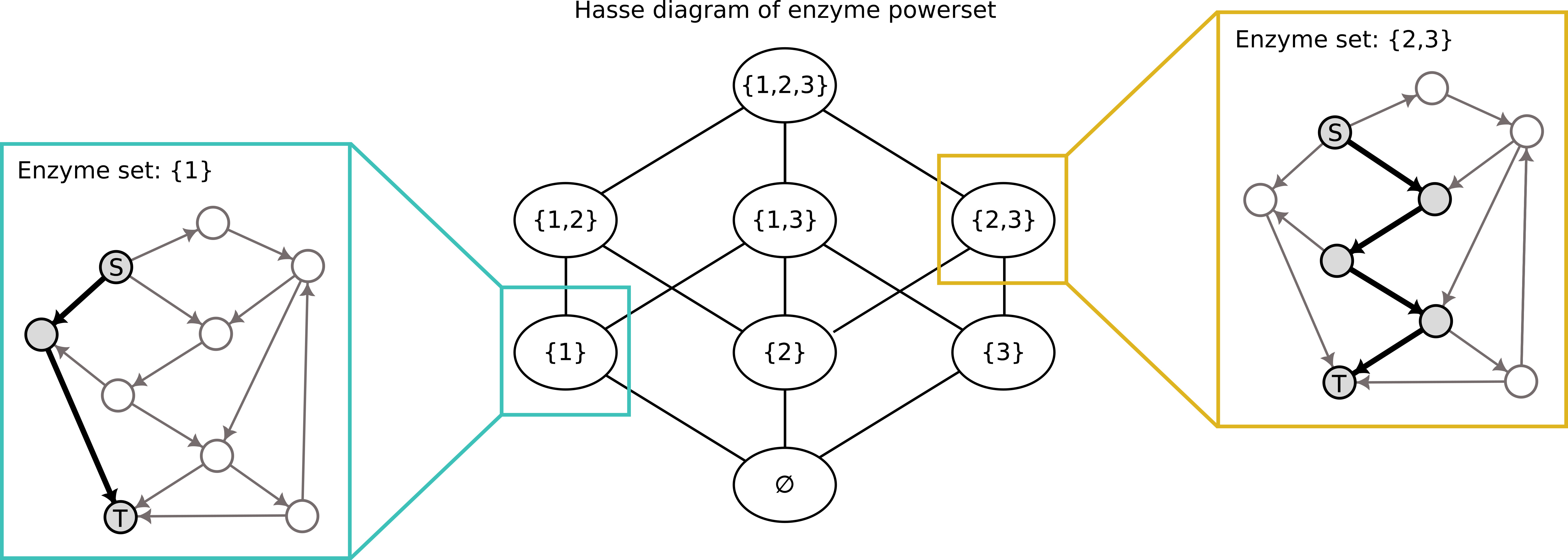}
    \label{fig:graph_abstract}
    \end{figure}

\end{abstract}

\section{Introduction}
A central promise of metabolic engineering and biomanufacturing is that useful molecules can be produced under milder conditions, with less waste, and with better use of carbon than in many conventional chemical processes \cite{planet_compatible_chemindustry_2023, kim_metabolic_2023, ko_tools_2020}. This promise is no longer limited to modifying whole cells. Cell-free systems also make it possible to assemble enzyme cascades directly and test pathway designs outside the constraints of a living host \cite{rasor_toward_2021}. As a result, pathway design is increasingly a question of choice: which enzymes should be included, which can be omitted, and which combinations are worth carrying forward experimentally?

This question matters because adding enzymes is not automatically beneficial. In \textit{in vivo} design, a pathway must operate in the context of the host metabolism, where additional reactions may interact with existing fluxes and side routes. In \textit{in vitro} design, every enzyme must be supplied, maintained, and made compatible with the rest of the cascade \cite{kim_metabolic_2023, rasor_toward_2021}. A smaller enzyme set may therefore be easier to implement, while a larger one may open alternative routes, introduce unnecessary complexity, or force metabolites through less efficient conversions.

Central carbohydrate metabolism illustrates this tension well. Glycolysis and the pentose phosphate pathway are often introduced as self-contained pathways, but biochemically they are embedded in a larger network of sugar-phosphate interconversions, branch points, and recycling reactions \cite{glycolysis_2026_gruening_review, PPP_review_stincone_2015}. Synthetic examples such as non-oxidative glycolysis (NOG) show that reorganizing familiar enzymatic chemistry can produce alternative carbon-conserving routes to valuable metabolites \cite{Bogorad_2013-nog, hellgren_promiscuous_2020}. More broadly, recent synthetic pathway studies suggest that useful designs can emerge from non-natural or reorganized enzyme combinations rather than from refining a single canonical route alone \cite{schwander_synthetic_2016, mclean_2023_hopac}.

The resulting design space is combinatorial. Once a biochemical system is described by a set of enzymes, the question of which enzymes are sufficient becomes a question over all enzyme subsets. Looking at one pathway, one deletion, or one candidate design at a time can miss the structure of this space: feasible regions may be sparse, some enzymes may be critical only in combination with others, and reduced subsets may retain high carbon efficiency.
In this paper, we make this subset structure of enzymes the object of analysis. 

Our analysis is computational and based on a formal modeling of enzymes and pathways. In particular, 
we model an enzyme by a graph transformation rule that defines how metabolites, represented as molecular graphs, might be transformed in a chemical reaction catalyzed by this enzyme. Such a rule can represent several reactions having the same core mechanism, and we assume that each rule employed represents all reactions that the actual enzyme may catalyze.
%
%
%
%
For every subset of enzyme rules of the given biochemical system, we then computationally generate the corresponding chemical reaction space and search for alternative pathways (i.e., pathways from prescribed input molecules to desired output molecules, as specified by the base pathway defining the system). In this process, we model reaction spaces as directed hypergraphs and pathways as integer hyperflows, following the framework of chemical transformation motifs \cite{flow}. 
The pathways returned by a search we rank by a metric based on pathway-level quantities such as carbon efficiency and pathway size.
Whenever some productive alternative pathway exists, we call the respective subset of enzymes feasible.

The feasible subsets of enzymes (rules)
are then organized by set inclusion and visualized as a Hasse diagram, where each subset of enzymes corresponds to a node. In this representation, moving upward corresponds to adding enzymes, while moving downward corresponds to removing them. Annotating each node with pathway metrics makes it possible to ask how feasibility and performance change across the subset landscape, and to identify dependencies, candidate redundancies, and small enzyme sets that still perform well.

We apply this approach to three systems: glycolysis in its Embden--Meyerhof--Parnas (EMP) form, the non-oxidative pentose phosphate pathway (non-oxPPP), and non-oxidative glycolysis (NOG) \cite{glycolysis_2026_gruening_review, PPP_review_stincone_2015, Bogorad_2013-nog}. These systems were chosen because they span core native metabolism, carbon rearrangement chemistry, and synthetic carbon-conserving pathway design. They therefore provide contrasting cases for asking how enzyme-subset structure shapes pathway feasibility and carbon efficiency.

Thus, the main contribution of this paper is a systematic subset-analysis approach for exploring biochemical pathway alternatives before committing to a single detailed design. The method is intended as a screening layer: it does not replace thermodynamic, kinetic, or experimental validation, but it can narrow attention to minimal working sets, structurally important enzymes, and reduced high-performing pathway candidates.

The remainder of the paper is structured as follows. Section~\ref{sec:preliminaries} covers necessary preliminaries and notation used in the paper. Section~\ref{sec:methods} goes into detail with the method developed in this paper. In Section~\ref{sec:system_specification} we present the specifications of the three systems that we consider. Lastly, in Section~\ref{sec:results} we present the results of applying the method to the three systems we are analyzing.

\section{Formal Modeling and Computational Tools}    
\label{sec:preliminaries}
\begin{figure}[hbt]
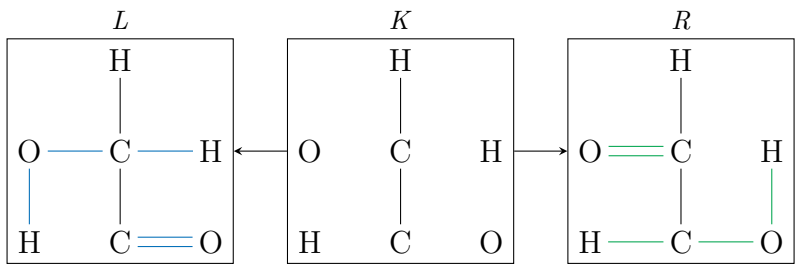

    \centering
    \dpoRule{tikz_figures/rule/001_r_0_10300000_L}{tikz_figures/rule/001_r_0_10300000_K}{tikz_figures/rule/001_r_0_10300000_R}
        
    \caption{Example of a graph transformation rule representing
    isomerase reactions
    from ketose to aldose. If a molecular graph contains the pattern $L$, then the blue part of the pattern will be replaced by the green part of the pattern $R$. The black part of the pattern is not changed, but must be present in the molecular graph for the rule to be applied.}
    \label{fig:rule}
\end{figure}

As evidenced by their widespread representation in textbooks, molecules are well modeled as undirected, labeled graphs, where atoms correspond to vertices and chemical bonds correspond to edges. Labels are used to describe atom type, bond order, and other types of information such as charge.
Based on this, chemical reactions can be modeled as \textit{graph transformation rules}~\cite{DBLP:conf/gg/CorradiniMREHL97,mod_paper}. A graph transformation rule describes how reactants (molecular graphs) are transformed into products (other molecular graphs). 
Each such rule consists of three graphs: $L, K$, and $R$. Here, $L$ is a pattern that must be present in the reactant graph---or a set of graphs, in the case of multiple reactants---in order to apply the rule. $R$ is the result (e.g., the restructuring of chemical bonds) of applying the rule at the position in the reactant(s) where the pattern~$L$ appeared. The graph $K$, often called the \textit{context}, describes what $L$ and $R$ have in common, i.e., what is part of~$L$ but not changed by the reaction. Fig.~\ref{fig:rule} illustrates a rule which represents an isomerase reaction between a ketose and an aldose sugar. The rule requires only the context of a ketone group, allowing the rule to apply to different ketose sugars instead of only one specific.

Given a starting set of molecules represented by molecular graphs and reactions represented by a set of graph transformation rules, we can use the software package~MØD~\cite{mod_paper, mod, flow} to \textit{in silico} expand a chemical space
by iteratively applying the graph transformation rules to the current set of molecules. Each step of the iteration may create new molecules, which are then added to the set of molecules. The process can be repeated until no new molecules appear or until a desired stopping criteria is met.

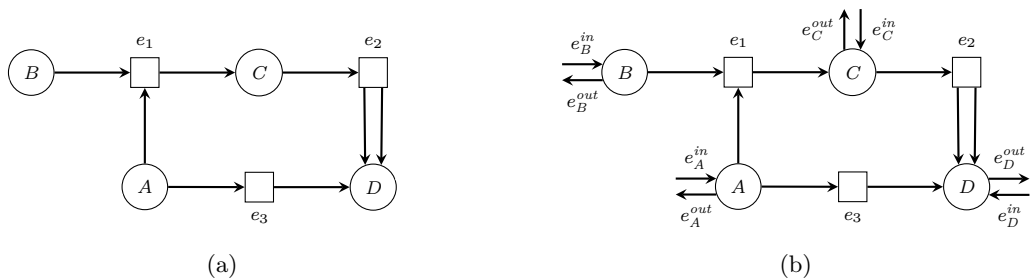
\begin{figure}[hbt]
\centering
\begin{subfigure}[t]{0.49\textwidth}
\centering
\scalebox{\scale}{
\begin{tikzpicture}[abstractNetwork]
    \node[vertex] (A) {$A$};    
    \node[vertexBase, overlay, left = 0.5 of A] (A1) {\phantom{$A$}};
    \node[hyperEdge, above = of A, label=above:$e_1$] (e1) {\phantom{$0$}};
    \node[vertex, left = of e1] (B) {$B$};
    \node[vertexBase, overlay, left = 0.5 of B] (B1) {\phantom{$B$}};
    \node[vertex, right = of e1] (C) {$C$};
    \node[vertexBase, overlay, above = 0.5 of C] (C1) {\phantom{$C$}};
    \node[hyperEdge, right = of C, label=above:$e_2$] (e2) {\phantom{$0$}};
    \node[vertex, below = of e2] (D) {$D$};
    \node[vertexBase, overlay, right = 0.5 of D] (D1) {\phantom{$D$}};
    \node[hyperEdge, left = of D, label=below:$e_3$] (e3) {\phantom{$0$}};

    \draw[arrow] (A) -- (e1);
    \draw[arrow] (B) -- (e1);
    \draw[arrow] (e1) -- (C);
    
    \draw[arrow] (C) -- (e2);
    \draw[arrow] (e2.300) -- (D.70);
    \draw[arrow] (e2.240) -- (D.110);

    \draw[arrow] (A) -- (e3);
    \draw[arrow] (e3) -- (D);

    \draw[arrow, white] (A.200) -- node[below] {$e_A^{\mathit{out}}$} (A1.340);
    \draw[arrow, white] (A1.20) -- node[above] {$e_A^{\mathit{in}}$} (A.160);

    \draw[arrow, white] (B.200) -- node[below] {$e_B^{\mathit{out}}$} (B1.340);
    \draw[arrow, white] (B1.20) -- node[above] {$e_B^{\mathit{in}}$} (B.160);

    \draw[arrow, white] (C.110) -- node[left] {$e_C^{\mathit{out}}$} (C1.250);
    \draw[arrow, white] (C1.290) -- node[right] {$e_C^{\mathit{in}}$} (C.70);

    \draw[arrow, white] (D.20) -- node[above] {$e_D^{\mathit{out}}$} (D1.340);
    \draw[arrow, white] (D1.160) -- node[below] {$e_D^{\mathit{in}}$} (D.20);
\end{tikzpicture}
}
\caption{}
\label{fig:hypergraph}
\end{subfigure}
\begin{subfigure}[t]{0.49\textwidth}
\centering
\scalebox{\scale}{
\begin{tikzpicture}[abstractNetwork]
    \node[vertex] (A) {$A$};    
    \node[vertexBase, overlay, left = 0.5 of A] (A1) {\phantom{$A$}};
    \node[hyperEdge, above = of A, label=above:$e_1$] (e1) {\phantom{$0$}};
    \node[vertex, left = of e1] (B) {$B$};
    \node[vertexBase, overlay, left = 0.5 of B] (B1) {\phantom{$B$}};
    \node[vertex, right = of e1] (C) {$C$};
    \node[vertexBase, overlay, above = 0.5 of C] (C1) {\phantom{$C$}};
    \node[hyperEdge, right = of C, label=above:$e_2$] (e2) {\phantom{$0$}};
    \node[vertex, below = of e2] (D) {$D$};
    \node[vertexBase, overlay, right = 0.5 of D] (D1) {\phantom{$D$}};
    \node[hyperEdge, left = of D, label=below:$e_3$] (e3) {\phantom{$0$}};

    \draw[arrow] (A) -- (e1);
    \draw[arrow] (B) -- (e1);
    \draw[arrow] (e1) -- (C);
    
    \draw[arrow] (C) -- (e2);
    \draw[arrow] (e2.300) -- (D.70);
    \draw[arrow] (e2.240) -- (D.110);

    \draw[arrow] (A) -- (e3);
    \draw[arrow] (e3) -- (D);

    \draw[arrow] (A.200) -- node[below] {$e_A^{\mathit{out}}$} (A1.340);
    \draw[arrow] (A1.20) -- node[above] {$e_A^{\mathit{in}}$} (A.160);

    \draw[arrow] (B.200) -- node[below] {$e_B^{\mathit{out}}$} (B1.340);
    \draw[arrow] (B1.20) -- node[above] {$e_B^{\mathit{in}}$} (B.160);

    \draw[arrow] (C.110) -- node[left] {$e_C^{\mathit{out}}$} (C1.250);
    \draw[arrow] (C1.290) -- node[right] {$e_C^{\mathit{in}}$} (C.70);

    \draw[arrow] (D.20) -- node[above] {$e_D^{\mathit{out}}$} (D1.160);
    \draw[arrow] (D1.200) -- node[below] {$e_D^{\mathit{in}}$} (D.340);

\end{tikzpicture}
}
\caption{}
\label{fig:extended_hypergraph}
\end{subfigure}
\caption{A simple CRN modeled as a directed multi-hypergraph $\mathcal{H}=(V,E)$ in \subref{fig:hypergraph} and the extended version of~$\mathcal{H}$ in \subref{fig:extended_hypergraph}.
Vertices are drawn as circles and hyperedges are drawn as squares.
The vertex set is $V = \{A, B, C, D\}$ and the set of hyperedges is $E = \{e_1, e_2, e_3\}$ in \subref{fig:hypergraph},
while \subref{fig:extended_hypergraph} has additional half-edges to and from every vertex.}
\end{figure}

We model such chemical spaces (either created by the iterative process above or by other means) as directed multi-hypergraphs, following the approach in~\cite{flow}. 
The vertices of the hypergraph represent the molecules of the space and the directed hyperedges, each an ordered pair of multisets of vertices, represent the reactions.
An example of such a directed multi-hypergraph is shown in Fig.~\ref{fig:hypergraph}. As a synonym for a chemical space modeled as a hypergraph, we will use the term \textit{chemical reaction network} (CRN).

To model \textit{pathways} in a CRN, we also follow ~\cite{flow}. To that end, a directed multi-hypergraph~$\mathcal{H} = (V, E)$ is extended with \textit{half-edges}. That is, for each vertex $v \in V$ we add two additional edges, one going from $\emptyset$ to $v$, denoted by $e^{\mathit{in}}_v$, and one going from $v$ to $\emptyset$, denoted by $e^{\mathit{out}}_v$. The resulting hypergraph is called the \textit{extended hypergraph} of~$\mathcal{H}$. Fig.~\ref{fig:extended_hypergraph} shows the extended hypergraph of the directed multi-hypergraph in Fig.~\ref{fig:hypergraph}.

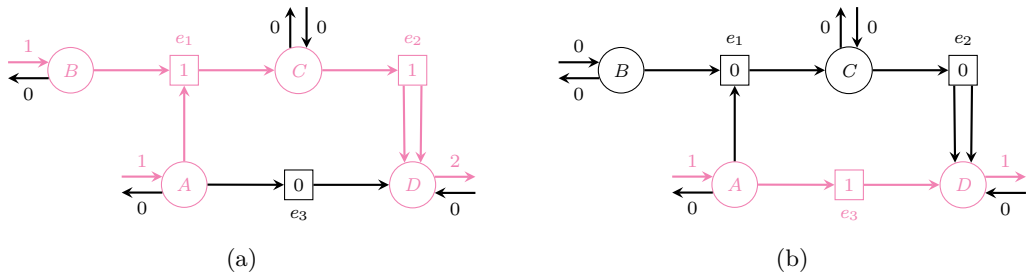
\begin{figure}[hbt]
\centering
\begin{subfigure}[t]{0.47\textwidth}
\centering
\scalebox{\scale}{
\begin{tikzpicture}[abstractNetwork]
    \node[vertex, \flowColor] (A) {$A$};    
    \node[vertexBase, overlay, left = 0.5 of A] (A1) {\phantom{$A$}};
    \node[hyperEdge, above = of A, \flowColor, label={[text=\flowColor] above:$e_1$}] (e1) {{$1$}};
    \node[vertex, left = of e1, \flowColor] (B) {$B$};
    \node[vertexBase, overlay, left = 0.5 of B] (B1) {\phantom{$B$}};
    \node[vertex, right = of e1, \flowColor] (C) {$C$};
    \node[vertexBase, overlay, above = 0.5 of C] (C1) {\phantom{$C$}};
    \node[hyperEdge, right = of C, \flowColor, label={[text=\flowColor]above:$e_2$}] (e2) {{$1$}};
    \node[vertex, below = of e2, \flowColor] (D) {$D$};
    \node[vertexBase, overlay, right = 0.5 of D] (D1) {\phantom{$D$}};
    \node[hyperEdge, left = of D, label=below:$e_3$] (e3) {{$0$}};

    \draw[arrow, \flowColor] (A) -- (e1);
    \draw[arrow, \flowColor] (B) -- (e1);
    \draw[arrow, \flowColor] (e1) -- (C);
    
    \draw[arrow, \flowColor] (C) -- (e2);
    \draw[arrow, \flowColor] (e2.300) -- (D.70);
    \draw[arrow, \flowColor] (e2.240) -- (D.110);

    \draw[arrow] (A) -- (e3);
    \draw[arrow] (e3) -- (D);

    \draw[arrow] (A.200) -- node[below] {$0$} (A1.340);
    \draw[arrow, \flowColor] (A1.20) -- node[above] {$1$} (A.160);

    \draw[arrow] (B.200) -- node[below] {$0$} (B1.340);
    \draw[arrow, \flowColor] (B1.20) -- node[above] {$1$} (B.160);

    \draw[arrow] (C.110) -- node[left] {$0$} (C1.250);
    \draw[arrow] (C1.290) -- node[right] {$0$} (C.70);

    \draw[arrow, \flowColor] (D.20) -- node[above] {$2$} (D1.160);
    \draw[arrow] (D1.200) -- node[below] {$0$} (D.340);
\end{tikzpicture}
}
\caption{}
\label{fig:flow1}
\end{subfigure}
\begin{subfigure}[t]{0.47\textwidth}
\centering
\scalebox{\scale}{
\begin{tikzpicture}[abstractNetwork]
    \node[vertex, \flowColor] (A) {$A$};    
    \node[vertexBase, overlay, left = 0.5 of A] (A1) {\phantom{$A$}};
    \node[hyperEdge, above = of A, label=above:$e_1$] (e1) {{$0$}};
    \node[vertex, left = of e1] (B) {$B$};
    \node[vertexBase, overlay, left = 0.5 of B] (B1) {\phantom{$B$}};
    \node[vertex, right = of e1] (C) {$C$};
    \node[vertexBase, overlay, above = 0.5 of C] (C1) {\phantom{$C$}};
    \node[hyperEdge, right = of C, label=above:$e_2$] (e2) {{$0$}};
    \node[vertex, below = of e2, \flowColor] (D) {$D$};
    \node[vertexBase, overlay, right = 0.5 of D] (D1) {\phantom{$D$}};
    \node[hyperEdge, left = of D, \flowColor, label={[text=\flowColor]below:$e_3$}] (e3) {{$1$}};

    \draw[arrow] (A) -- (e1);
    \draw[arrow] (B) -- (e1);
    \draw[arrow] (e1) -- (C);
    
    \draw[arrow] (C) -- (e2);
    \draw[arrow] (e2.300) -- (D.70);
    \draw[arrow] (e2.240) -- (D.110);

    \draw[arrow, \flowColor] (A) -- (e3);
    \draw[arrow, \flowColor] (e3) -- (D);

    \draw[arrow] (A.200) -- node[below] {$0$} (A1.340);
    \draw[arrow, \flowColor] (A1.20) -- node[above] {$1$} (A.160);

    \draw[arrow] (B.200) -- node[below] {$0$} (B1.340);
    \draw[arrow] (B1.20) -- node[above] {$0$} (B.160);

    \draw[arrow] (C.110) -- node[left] {$0$} (C1.250);
    \draw[arrow] (C1.290) -- node[right] {$0$} (C.70);

    \draw[arrow, \flowColor] (D.20) -- node[above] {$1$} (D1.160);
    \draw[arrow] (D1.200) -- node[below] {$0$} (D.340);
\end{tikzpicture}
}
\caption{}
\label{fig:flow2}
\end{subfigure}
\caption{Two integer hyperflows on the extended hypergraph from Fig.~\ref{fig:extended_hypergraph}. The value that the flow associates with each edge is written in the square of the hyperedge. For the half-edges, the value is written next to it. The hyperedges with positive flow and the vertices that are endpoints of such an edge are coloured pink. In~(a), the set of input molecules is~$\{A,B\}$ and the set of output molecules is~$\{D\}$, while in~(b), they are $\{A\}$ and~$\{D\}$, respectively. 
}
\label{fig:flows}
\end{figure}

A pathway in a CRN (a hypergraph) can then be modeled~\cite{flow} by an \textit{integer hyperflow}~$f$ on the extended hypergraph by associating each edge~$e$ with a non-negative integer number~$f(e)$. This number specifies how many times the reaction that the edge~$e$ models occurs in the pathway~\cite{flow}.
Additionally, as usual in metabolic control analysis
we require a flow to be stationary, that is,
for all~$v$, the total flow into~$v$ (the production of $v$ from $f(e)$ instances of $e$, summed over all $e$ in the extended hypergraph) is equal to the total flow leaving~$v$ (the consumption of $v$ from $f(e)$ instances of $e$, summed over all $e$ in the extended hypergraph). 
%
If a vertex $v\in V$ is an input to the pathway, i.e. $f(e_v^{\mathit{in}})>0$, we call it an \textit{inflow molecule}, and if it is an output of the pathway, i.e. $f(e_v^{\mathit{out}})>0$, we call it an \textit{outflow molecule}. Examples of two integer hyperflows are given in Fig.~\ref{fig:flows}.

Using MØD, we can search for integer hyperflows (pathways) that fulfil a given specification~\cite{mod, flow}. Such a search is called a \textit{flow query}. 
A flow query can, for example, specify the inflow molecule(s) of the pathway and the outflow molecule(s). Consider again the flows in Fig.~\ref{fig:flows}. Here both flows fulfil the flow query that $A$ must be an inflow molecule, that $B$ may be an inflow molecule, and that $D$ must be an outflow molecule. When several flows fulfil a flow query, they can be ranked by defining an \textit{objective function}~\cite{mod, flow}. An objective function can for instance maximize the amount of product produced or minimize the number of unique internal edges used. In our examples in~Fig.~\ref{fig:flows}, the flow in~\subref{fig:flow1} produces more of $D$, but the flow in~\subref{fig:flow2} uses fewer internal edges and the sum of the flow on the internal edges is lower.

\section{Methodology}
\label{sec:methods}


The key modeling distinction in this paper is not whether a pathway is literally implemented \textit{in vivo} or \textit{in vitro}, but how much chemical context is made available before the subset-specific pathway search is performed. In the full-set-generated space (FGS) setting, the chemical reaction network is expanded once from the complete enzyme set $S$. Each subset $s \subseteq S$ is then evaluated inside this common background space, while the pathway itself is still required to use every enzyme in $s$ at least once. FGS therefore asks whether a subset can form a productive pathway when intermediates generated by the larger system may be available. In the subset-generated space (SGS) setting, the chemical reaction network is generated separately for each subset $s$. SGS therefore asks whether the subset can both generate the necessary intermediates and carry the productive pathway on its own. The two settings should be interpreted as complementary structural assumptions: FGS tests subset performance in a chemically richer background, whereas SGS tests self-contained subset performance.    

\begin{figure}[htbp]
\centering
    \begin{subfigure}[t]{0.48\textwidth}
        \centering
        \begin{tikzpicture}[font=\scriptsize, node distance = 0.5cm]
        
        \node (start) [startstop] {Start};
        \node (pro1) [process, below = of start] {Select next set $s \in \mathcal{P}(S)$};
        \node (pro2) [process, below =of pro1] {Generate the CRN spanned by $s$};
        
        \node (pro3) [process, below =of pro2] {Search for pathways spanned by the entire set $s$ in the CRN};
        
        \node (dec0) [decision, below = of pro3, align=center] {More\\subsets to\\process?};
        \node (stop) [startstop, below = of dec0] {Stop};
        
        \draw [arrow] (start) -- (pro1);
        \draw [arrow] (pro1) -- (pro2);
        \draw [arrow] (pro2) -- (pro3);
        \draw [arrow] (pro3) -- (dec0);
        \draw [arrow] (dec0.east) -| ++ (1,0) |-  (pro1);
        \coordinate (rightshift) at ($(dec0.east)-(dec0.center) + (1.3,0)$);
        \coordinate(downshift) at ($(pro1)!0.5!(dec0)$);
        \node[] at ($(downshift)+(rightshift)$) {Yes};
        \draw [arrow] (dec0) -- node[midway, xshift=0.3cm] {No} (stop);
        \end{tikzpicture}

        \caption{SGS.}
        \label{fig:flowchart_subspace}
    \end{subfigure}
    \begin{subfigure}[t]{0.48\textwidth}
        \centering
        \begin{tikzpicture}[font=\scriptsize, node distance = 0.5cm]
        
        \node (start) [startstop] {Start};
        \node (pro1) [process, below =of start] {Generate the CRN spanned by $S$};
        
        \node (pro2) [process, below = of pro1] {Select next set $s \in \mathcal{P}(S)$};        
        \node (pro3) [process, below =of pro2] {Search for pathways spanned by the entire set $s$ in the CRN};
        
        \node (dec0) [decision, below = of pro3, align=center] {More\\subsets to\\process?};
        \node (stop) [startstop, below = of dec0] {Stop};
        
        \draw [arrow] (start) -- (pro1);
        \draw [arrow] (pro1) -- (pro2);
        \draw [arrow] (pro2) -- (pro3);
        \draw [arrow] (pro3) -- (dec0);
        \draw [arrow] (dec0.east) -| ++ (1,0) |-  (pro2);
        \coordinate (rightshift) at ($(dec0.east)-(dec0.center) + (1.3,0)$);
        \coordinate(downshift) at ($(pro2)!0.5!(dec0)$);
        \node[] at ($(rightshift)+(downshift)$) {Yes};
        \draw [arrow] (dec0) -- node[midway, xshift=0.3cm] {No} (stop);
        \end{tikzpicture}
        \caption{FGS.}
        \label{fig:flowchart_wholespace}
    \end{subfigure}

    \caption{Flowcharts representing the SGS method in \subref{fig:flowchart_subspace} and the FGS method in \subref{fig:flowchart_wholespace}. $S$ is the set of transformation rules that model the enzymatic reactions of the system to be explored and $\mathcal{P}(S)$ denotes the powerset of $S$. Remark that the pathways that are found must use each enzyme in the set $s$ at least once.}
    \label{fig:flowchart}
\end{figure}
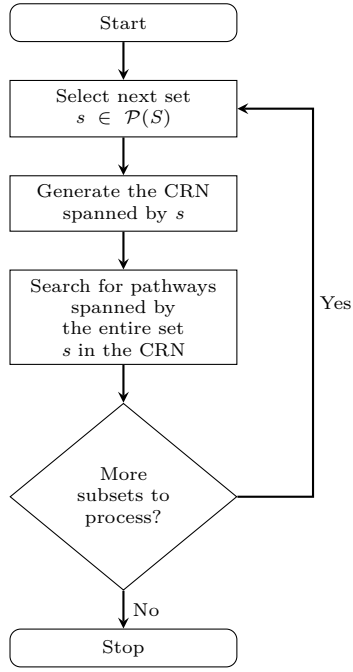
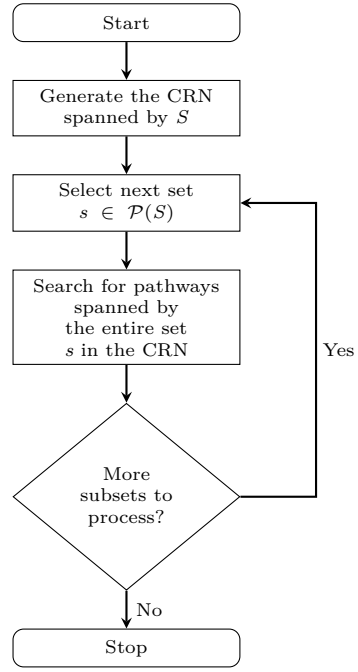

We present the workflow of the methods in Fig.~\ref{fig:flowchart}. Both methods use a set of graph transformation rules (for short \textit{rules}) that model the enzymatic reactions known to the system, denoted by $S$ and the powerset of $S$ denoted by $\mathcal{P}(S)$.

    The SGS method starts by choosing a set $s\in \mathcal{P}(S)$, then it expands a CRN from $s$ and the starting molecules of the system. Subsequently, it searches for pathways. These pathways convert a set of reactants to a specified product while using all rules in $s$ at least once. Lastly, another set $s' \in \mathcal{P}(S)$ is chosen and the process is repeated. We enforce the use of all enzymes in the subset to ensure the discovery of novel solutions instead of carrying over legacy solutions from subset to superset.
    
    The FGS method, on the other hand, generates one CRN from the entire set of rules $S$ and the starting molecules. In this CRN, it searches for pathways for each set $s\in \mathcal{P}(S)$ identical to how it is done in the SGS method. That is, the two methods only differ in how the CRN is expanded.

    As mentioned, the pathways are required to convert a set of reactants (inflow molecules) to the product (a required outflow molecule). Further, we allow all other molecules to be outflow molecules. However, a molecule cannot both be an inflow- and an outflow molecule of the pathway. 
    For the pathways that fulfill these requirements, we are interested in the ones that firstly maximizes the amount of the product molecule $P$, while subsequently minimizing the number of unique edges with a strictly positive flow denoted by $|E_{>0}|$ and lastly minimizes the amount of flow in the pathway. We do this by a hierarchical objective function. That is, when we search for pathways, we discard all those that do not give an optimal value for the function

    \begin{equation}
        \min{\bigl(-10000\cdot f(e^{\mathit{out}}_P) + 100 \cdot |E_{>0}| + \sum_{e\in E} f(e)\bigr)}.
    \end{equation}

    \noindent If there exists more than 100 optimal flows for a flow query, we only consider the first 100 of them.

\section{Biochemical Systems Description}\label{sec:system_specification}
    We here specify the modeling of the three systems non-oxPPP, NOG and Glycolysis.
    In Table~\ref{tab:enz_names} we present the enzymes used for each system and their abbreviations. The rules to model the enzymes in the systems can be found in the supplementary material and were modeled according to the approaches in~\cite{flow} and~\cite{50_shades_andersen}. These rules represent the core of the reaction that is catalyzed, with the necessary context and restrictions to stay close to the enzymatic reaction with a margin for promiscuity. Further information on the imposed constraints for the rules can be found in the appendix.
    
    In Table~\ref{tab:exp_setup} we present the specific settings used for each chemical system for generating its CRN and enumerating the pathways within. Note that for some reactants we specify the amount required in the pathway, e.g. for non-oxPPP we require an inflow of 6 Ru5P. If no amount is specified, any amount of the reactant is allowed, e.g. the amount of water molecules is not enforced, therefore there can be any non-negative number of water molecules. 

    The settings for the expansion of the CRNs and the pathway finding lead to some modeling choices for each of the three systems. For non-oxPPP and Glycolysis, we have chosen that the starting molecules are those typically considered the starting materials or necessary cofactors of the pathway in question, as detailed in Table~\ref{tab:exp_setup}. Both pathways are modeled with the naturally occurring in- and outflows. The modeling choices for NOG are taken to follow the approach from \cite{flow}.  

    
\begin{table}
    \caption{Abbreviations of rule names with the reaction they were modelled after, including a short description of the functionality.}
    \label{tab:enz_names}
    \centering
    \begin{tabular}{lp{5cm}p{7cm}}
    \toprule
        \multicolumn{2}{c}{\textbf{non-oxPPP}}\\ 
        \midrule
        AL & Aldolase & Generic aldol addition \\
        AlKe B & Ketose-Aldose (backwards) & Ketone to aldehyde conversion \\
        AlKe F & Aldose-Ketose (forwards) & Aldehyde to ketone conversion \\
        TAL & Transaldolase & Move C3 unit \\
        TKL & Transketolase & Move C2 unit \\
        PHL & Phosphohydrolase & Use water to cleave off phosphate\\
        \midrule

        \multicolumn{2}{c}{\textbf{NOG} (same enzymes as \textbf{non-oxPPP} plus)} \\
        \midrule
        PK & Phosphoketolase & Break C-C-bond and add phosphate \\
        \midrule

        \multicolumn{2}{c}{\textbf{Glycolysis}} \\
        \midrule
        AL R & Aldolase (backwards) & Generic aldol cleavage \\
        AlKe B & Ketose-Aldose (backwards) & Ketone to aldehyde conversion \\
        EL & Enolase & Splitting off of alcohol group as water\\
        FI & Furanose & Furanose ring opening \\
        KEI & Keto-Enol Isomerase & Isomerisation from ketone to enol \\
        PDH & Phosphate Dehydrogenase & NAD-dependet bis-phosphorylation\\
        PGI & Pyranose-Furanose Isomerase & Isomerisation from C6 to C5 ring \\
        PK B & ATP Kinase (backward reaction) & Dephosphorylation and ATP generation \\        
        PK F & ATP Kinase (forward reaction) & Phosphorylation of an alcohol group \\
        PM & Phosphomutase & Intramolecular transfer of a phosphate \\        
    \bottomrule
    \end{tabular}
    \end{table}

\begin{table}
    \caption{Modeling setup for each individual system used in the analysis. The \textit{starting molecules} seed the chemical-space expansion; the \textit{rules} are the graph transformation rules used for expansion; pathway inputs and required products define the flow query. For PPP and Glycolysis the modeling followed the natural set up of the respective pathway with regards to starting molecules and inflow and outflow molecules. NOG was modeled following the approach described in \cite{flow}.}
    \label{tab:exp_setup}
    \centering
    \small
    \renewcommand{\arraystretch}{1.22}
    \setlength{\tabcolsep}{5pt}
    \begin{tabular}{@{}>{\raggedright\arraybackslash}p{0.18\textwidth}>{\raggedright\arraybackslash}p{0.24\textwidth}>{\raggedright\arraybackslash}p{0.27\textwidth}>{\raggedright\arraybackslash}p{0.23\textwidth}@{}}
        \toprule
        \rowcolor{Gray!18}
        \textbf{Model component} & \textbf{non-oxPPP} & \textbf{NOG} & \textbf{Glycolysis}\\
        \midrule
        \rowcolor{Gray!8}
        Starting molecules
        for expansion
        & Ru5P, water
        & AcP, G3P, DHAP, Pi, E4P, R5P, F6P, Ru5P, FBP, S7P
        & ATP, NAD\textsuperscript{+}, glucose, Pi, H\textsuperscript{+}, water\\
        \midrule
        Rules
        & AL, PHL, AlKe B, TAL, AlKe F, TKL
        & AL, PHL, AlKe B, TAL, AlKe F, TKL, PK
        & AL R, PDH, AlKe B, PGI, EL, PK B, FI, PK F, KEI, PM\\
        \midrule
        \rowcolor{Gray!8}
        Required inputs
        & 6 Ru5P
        & 1 F6P
        & 1 glucose\\
        \midrule
        Optional inputs & water & Pi, water & NAD\textsuperscript{+}, ADP, NADH, ATP, Pi, H\textsuperscript{+}, water\\
        \midrule
        \rowcolor{Gray!8}
        Required product
        & $\geq 1$ F6P
        & $\geq 1$ AcP
        & $\geq 1$ pyruvate\\
        \bottomrule
    \end{tabular}
\end{table}

\begin{figure}
\centering
\scalebox{0.29}{
\renewcommand\modDGHyperScale{1}
  \renewcommand\modInputPrefix{tikz_figures/}%
  \input{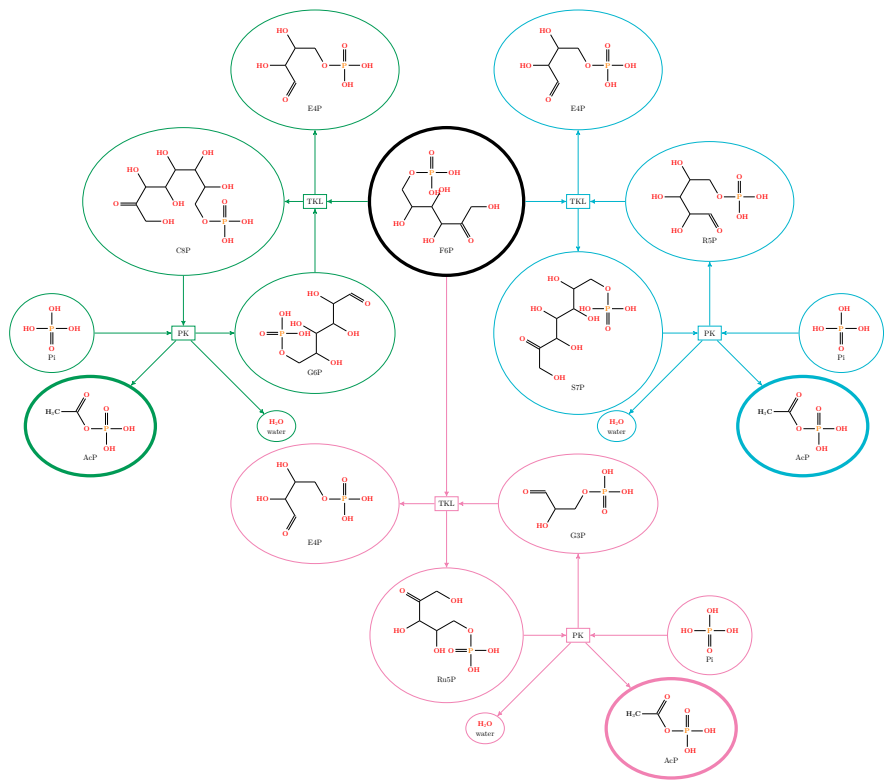}%
}
\caption{An example of three different pathways in NOG that only use the enzymes PK and TKL. Each pathway is coloured a different colour (green, pink and turquoise). We omit half-edges and flow numbers. All pathways start in F6P, as it is a required inflow molecule. They all end in AcP, as it is the product of the pathways. In addition, all pathways use Pi as an inflow molecule, and water and E4P as outflow molecules. Molecules (except F6P) that are repeated across pathways have been duplicated to avoid clutter.
The pink and turquoise pathways can be found both by the SGS method and the FGS method. The green pathway, on the other hand, can only be found using the FGS method. This is because G6P and C8P cannot be created from the starting molecules by only using the enzymes PK and TKL. Additional enzymes are needed to ensure that these molecules are present. When they are present in the system already (as in the FGS method), the pathway can run.}
\label{fig:min_ex}
\end{figure}

Now with the understanding of the biochemical systems modeling, we introduce an example to illustrate the influence of the FGS and the SGS settings on the outcome of the modeling.
Consider Fig.~\ref{fig:min_ex}, where we present three pathways found for NOG. Here we are working with the enzyme set $S=\{\text{AlKe F, AlKe B, TAL, TKL, PHL,}$ $\text{AL, PK}\}$ and have chosen the subset $s = \{\text{PK, TKL}\}$ as the current set the methods are working on. The chemical reaction network for the FGS method has been generated using all enzymes in $S$ and a smaller CRN has been generated for the SGS method with $s$. 
When we search for pathways spanned by $s$ in the two networks we find the pink and turquoise pathways with both methods and the green one only with the FGS method. The reason that the green pathway cannot be found using the SGS method, is that G6P and C8P cannot be created only using the enzymes TKL and PK and the starting molecules (see Table~\ref{tab:exp_setup}) but if they are already present in the system (as they are in the FGS method) they can be utilized in the pathway.

\section{Results and Discussion}
\label{sec:results}    
The following results should be read as an investigation of enzyme subset landscapes rather than as isolated pathway searches. For each system, the Hasse diagram records which enzyme subsets admit at least one pathway under the chosen modeling assumptions, and the node annotations summarize the best pathways found for those subsets. 
This makes three types of structure visible at once: the sparsity of the feasible region, the dependence of feasibility on particular enzyme combinations, and the extent to which carbon efficiency and optimal pathway size change when enzymes are added or removed. Carbon efficiency in this context relates to the ratio of output carbons to input carbons. The central question is therefore not only whether a pathway exists, but how robustly productive behavior is distributed across the enzyme-subset space.

Inside each node in the Hasse diagram, we indicate which enzymes are included in the subset, the number of optimal pathways found for that subset, and the number of reactions used by the pathways shown. 
All pathways for the same subset are identical in terms of product produced, unique edges used and sum of flow through the pathway. However, they may differ in which intermediates they produce as well as which reactions are used.

    
\subsection{The Non-Oxidative Pentose Phosphate Pathway}
\label{sec:results_PPP}
\begin{figure}
    \centering
    \includegraphics[width=\textwidth, clip, trim=6.5cm 2cm 6.5cm 0]{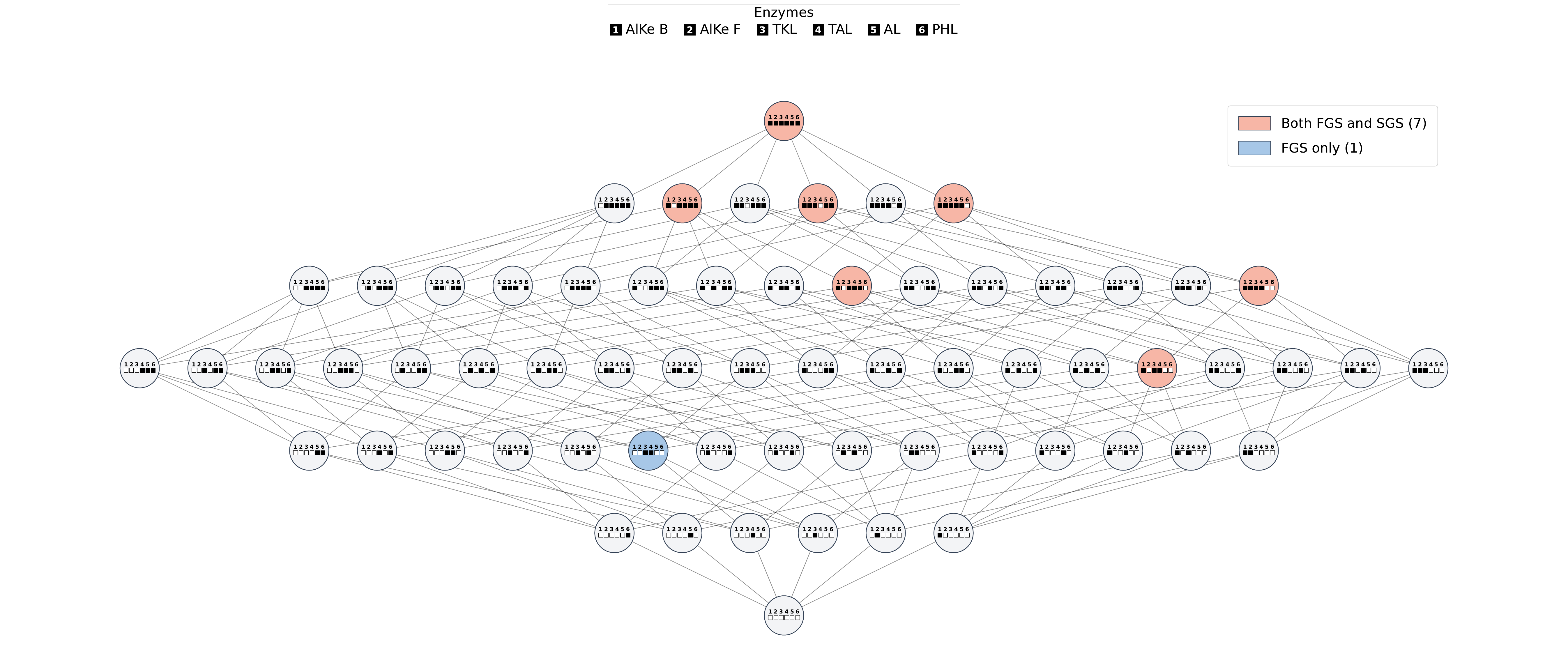}
    \caption{Hasse Diagram showing all possible enzyme subsets for non-oxPPP. Each node represents an enzyme subset. The enzymes which are present in the subset are in black in the vector of the node. A node is highlighted if a pathway can be found using exactly the enzymes in the subset at least once. The node is highlighted in blue if the pathway can only be found using the FGS method. The node is highlighted in coral if the pathway can be found by both the FGS and the SGS method.}
    \label{fig:PPP_feasible_region}
\end{figure}

    Figure~\ref{fig:PPP_feasible_region} shows the Hasse diagram of the full powerset for non-oxPPP. The nodes that admit a pathway under each of the schemes, are highlighted. We remind the reader that each rule in the set must be used at least once. Note that, if this was not a requirement, a set of enzymes which is a superset of another, would contain the optimal pathways of all its subsets. Therefore it would not be possible to tell if an additional enzyme might degrade the quality of a solution.

    In total, $8$ out of $2^{6}=64$ subsets admit a pathway, which corresponds to $12.5\%$ of the powerset. 
    These subsets are scattered, but most of them appear near the upper levels of the diagram, where more enzymes are present. The smallest subset contains only two enzymes, TKL and TAL (enzyme numbers 3 and 4), but this subset only allows a pathway in the FGS setting, because the pathway requires molecules from a catalytic cycle that cannot be generated from TKL and TAL alone.
    
    Another interesting finding is that across these subsets, the enzyme TKL is always present, which suggests that this is a necessary enzyme for our model to produce the target molecule.
    
\begin{figure}
    \centering
    \includegraphics[width=0.6\textwidth]{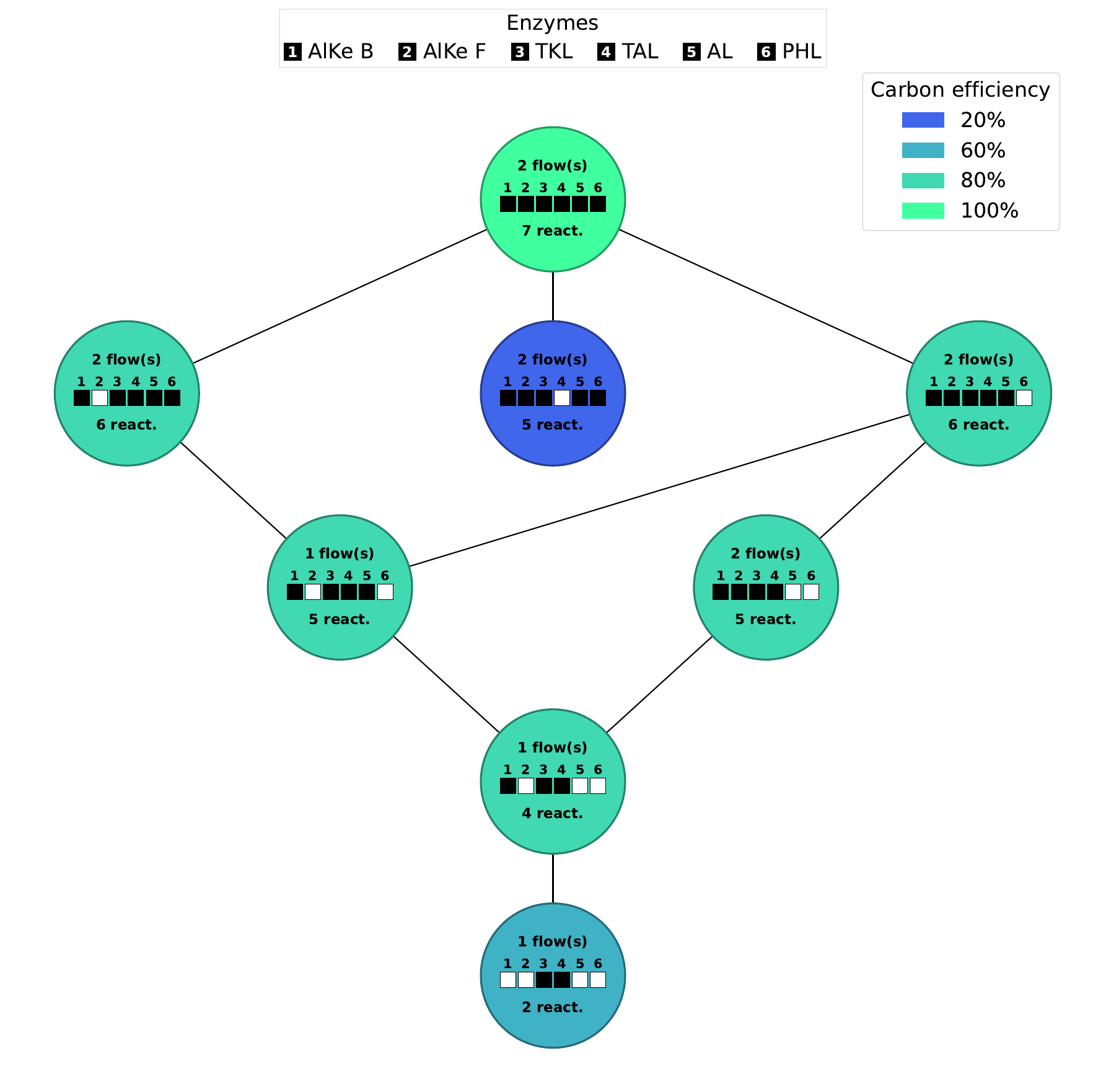}
    \caption{The Hasse diagram showing the feasible enzyme subsets of applying the FGS method to non-oxPPP. The Hasse diagram for the SGS method is identical, except that there is no pathway using the enzymes TKL and TAL (enzyme numbers 3 and 4).}
    \label{fig:PPP_whole_space}
\end{figure}

    We now focus on the subsets for which we found a pathway and analyze their carbon efficiency values. 
    Figure~\ref{fig:PPP_whole_space} shows the Hasse diagram for FGS. SGS is identical, except it cannot produce the pathway with the enzymes TKL and TAL (enzymes number 3 and 4).   
    Non-oxPPP can produce at most five F6P molecules in our query. Since we input 30 carbons, the possible carbon efficiency values are $\left\{ \frac{k}{5} \,\middle|\, k \in \{1,2,3,4,5\} \right\}$. In both settings, the carbon efficiency ranges from $20\%$ to $100\%$. Only one subset performs optimally: the full set of enzymes located at the root node of the diagram. Notice that the biggest difference in carbon efficiency is seen between the pathways that use all enzymes and have $100\%$ carbon efficiency, and the pathways that use all enzymes but TAL (enzyme number 4), with a contrasting value of $20\%$. 
    Notably, all other pathways use this enzyme and have a carbon efficiency of at least $60\%$.
    This could be due to TAL's ability to move C2 units, which is essential to the rearrangement reactions of non-oxPPP.

    Regarding the subset that contains only the enzymes TKL and TAL and is only productive in the FGS setting, the reason for this behavior is a 
    reaction that strictly reshuffles molecules, and therefore requires all carbons that ultimately end up in the product to already be present. Put simply, the functional group is getting shifted around. This is not possible in the SGS setup, since the molecules need to be produced from the subset in this setting in order to partake in the reaction. When these necessary molecules are available, the non-oxPPP target can be produced using reactions enabled by only these two enzymes, in contrast with the original size of six enzymes. 

\subsection{Non-Oxidative Glycolysis}
\label{sec:results_NOG}
We first identify the nodes within the Hasse diagram of the full powerset that admit a pathway under each of the schemes for NOG. Figure~\ref{fig:NOG_feasible_region} in the Appendix shows which enzyme subsets admit at least one valid pathway producing AcP under the NOG query.

In total, $25$ out of $2^{7}=128$ subsets admit a pathway, and these are distributed across all the levels of the diagram (disregarding, naturally, the empty set). Across these subsets, the PK enzyme (number $7$) appears in every pathway, which indicates that this enzyme is necessary for feasibility in our model.

\begin{figure}
    \centering
    \includegraphics[width=0.92\textwidth]{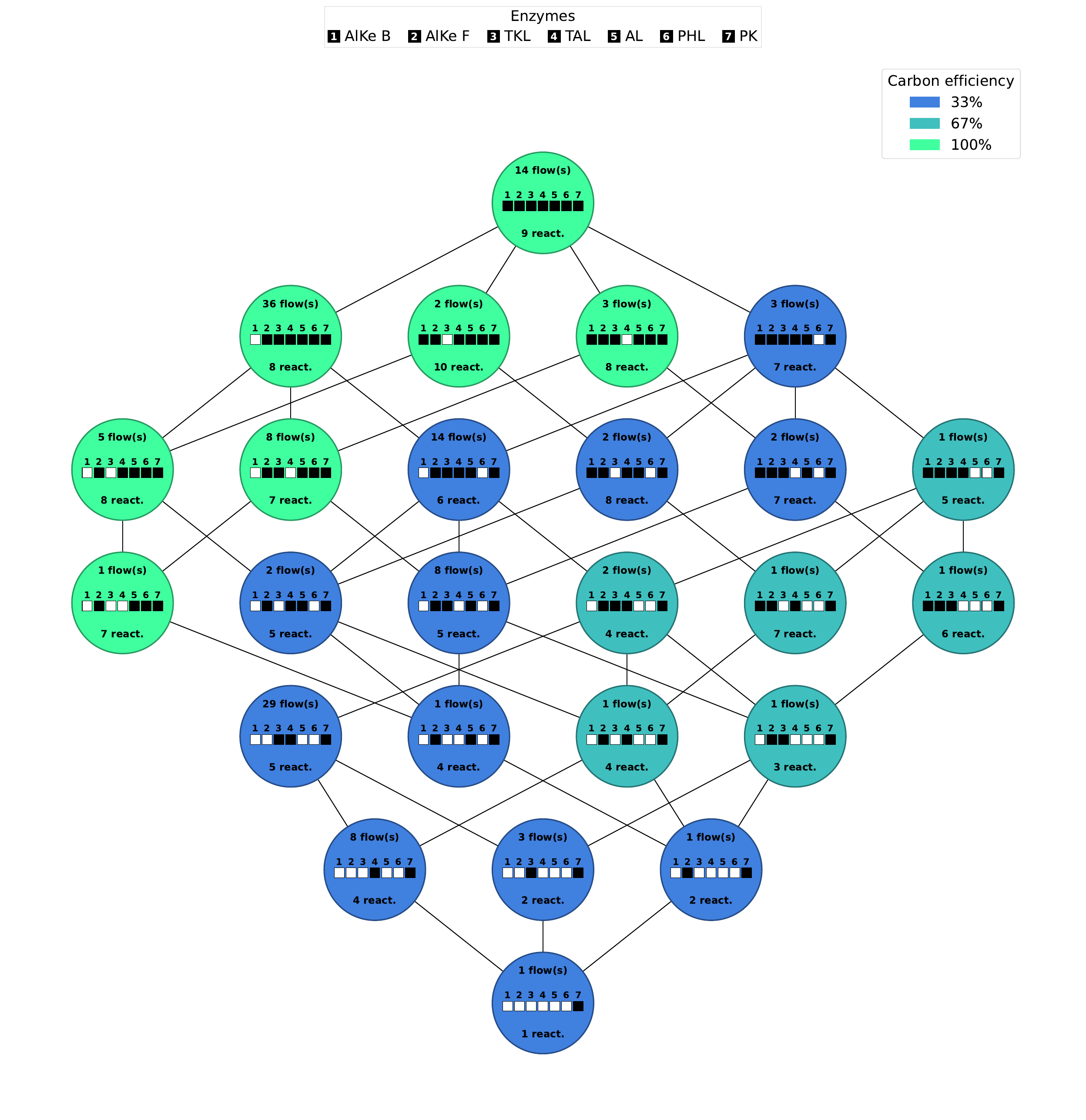}

    \caption{The Hasse diagram showing the results of applying the FGS method to NOG. We note that the Hasse diagram for the SGS method is identical, except that the node representing the enzyme set $\{\text{TKL}, \text{PK}\}$ (enzyme numbers 3 and 7) produces two solutions in the SGS method instead of three in the FGS method. 
    The ancestors of some nodes have a lower carbon efficiency than some of their descendants. This is because pathways are enforced to use each enzyme in the subset at least once. Three regions for carbon efficiency are visible: The $\sim33\%$ efficient enzyme sets up to level 2; the $\sim67\%$ efficient sets on the right side; and the 100\% efficient sets that include a set as small as 4 enzymes that is still able to convert $100\%$ of input carbons into product carbons. This set represents an alternative solution to NOG as described by Andersen and colleagues \cite{flow}, and later implemented by Hellgren and colleagues \cite{hellgren_promiscuous_2020}.} 
    \label{fig:NOG_wholespace}
\end{figure}

Next, we compare the subsets by carbon efficiency. The flow query for NOG can produce at most three AcP molecules. Since we input six carbon atoms, the possible carbon efficiencies are $\left\{ \frac{k}{3} \,\middle|\, k \in \{1,2,3\} \right\}$. Figure~\ref{fig:NOG_wholespace} shows the Hasse diagram for the FGS results. The Hasse diagram corresponding to the SGS setting differs from Fig.~\ref{fig:NOG_wholespace} only in that the subset containing enzymes TKL and PK ($3$ and $7$ respectively) yields three pathways for the FGS setup, and two for SGS.

NOG contains three clusters of nodes: one with $100\%$ carbon efficiency, one with $\sim67\%$ carbon efficiency, and one with $\sim33\%$ carbon efficiency. The $100\%$ carbon efficiency cluster (the light green nodes) shows that NOG in our model, in contrast to non-oxPPP, has several pathways that are optimal in carbon efficiency. Each of them represents reduced enzyme sets in which all the input carbons can be recovered in the target product. The light green nodes are therefore candidate pathways for shorter versions of our NOG model. One of the pathways uses the subset $\{\text{AlKe F, AL, PHL, PK}\}$ (enzyme numbers 2, 5, 6, and 7). A theoretical pathway using this subset of enzymes was first proposed by Andersen et al.~\cite{flow}, 
and was subsequently implemented experimentally by Hellgren and colleagues under the name Glycolysis AlTernative High Carbon Yield Cycle (GATHCYC)~\cite{hellgren_promiscuous_2020}. The $\sim67\%$ carbon efficiency cluster (in teal) is remarkably not connected to the $100\%$ cluster. It is surrounded by nodes with a carbon efficiency of $\sim33\%$.

Lastly, the $\sim33\%$ carbon efficiency cluster (in blue), notably, contains the pathway with an enzyme subset of size $1$, an enzyme subset of size $6$, and every size in between. This supports the hypothesis that it is not the size of the subset that has the greatest effect on efficiency, but the specific combination of enzymes in the subset.

\subsection{Glycolysis}
\label{sec:results_glycolysis}

    First, we identify the subsets within the Hasse diagram of the full powerset that admit a pathway under the FGS and SGS methods. Fig.~\ref{fig:Glycolysis_full_lattice} in the Appendix shows which enzyme subsets admit at least one pathway producing pyruvate, under our Glycolysis query. Levels without highlighted nodes have been left out for readability.

    The Glycolysis model contains $10$ enzyme rules, therefore the full powerset contains $2^{10}=1024$ subsets. Our results show that only four subsets admit a pathway in both SGS and FGS settings, corresponding to approximately $0.39\%$ of the powerset. Compared to non-oxPPP and NOG, Glycolysis subsets form a very small part of the combinatorial space. These nodes appear near the upper levels of the diagram, which means that the production of pyruvate for our query requires most of the enzymes. This difference between non-oxPPP and NOG versus Glycolysis could be explained by the linear structure of Glycolysis. While NOG and non-oxPPP have a more circular, recombinatorial structure, Glycolysis in nature is a more linear degradation of glucose. It is therefore notable that any subsets with productive pathways could be found at all.

    In all considered subsets, all enzymes apart from AlKe B and PM (enzyme numbers $3$ and $7$) are present. This means that these two enzymes are not necessary for Glycolysis to function in our model.
    Alke B represents the ketose-aldose reaction, catalyzing the transformation of a ketose sugar to an aldose one. Since we create a large combinatorial space before performing the flow queries, it is possible that the necessary intermediates were produced via a different route, and therefore, the Alke B catalyzed reaction is not necessary for a productive pathway. Note that this is not necessarily due to alternative enzymes making alternative routes possible, but due to alternative products of the enzymes included in the subset that enable product production through promiscuity. 
    A similar logic can be followed for PM, which represents the phosphomutase reaction. Here, the enzyme catalyzes the intramolecular transfer of a phosphate group from one hydroxyl group to another.

    Moving from the overall possibility of a node to produce a pathway to the quality of the pathways, we next compare the Glycolysis subsets by carbon efficiency. Our model can produce at most two pyruvate molecules. Since we input six carbons, the possible carbon efficiencies are $\left\{ \frac{k}{2} \,\middle|\, k \in \{1,2\} \right\}$.

    \begin{figure}
    \centering
    \includegraphics[scale=0.3]{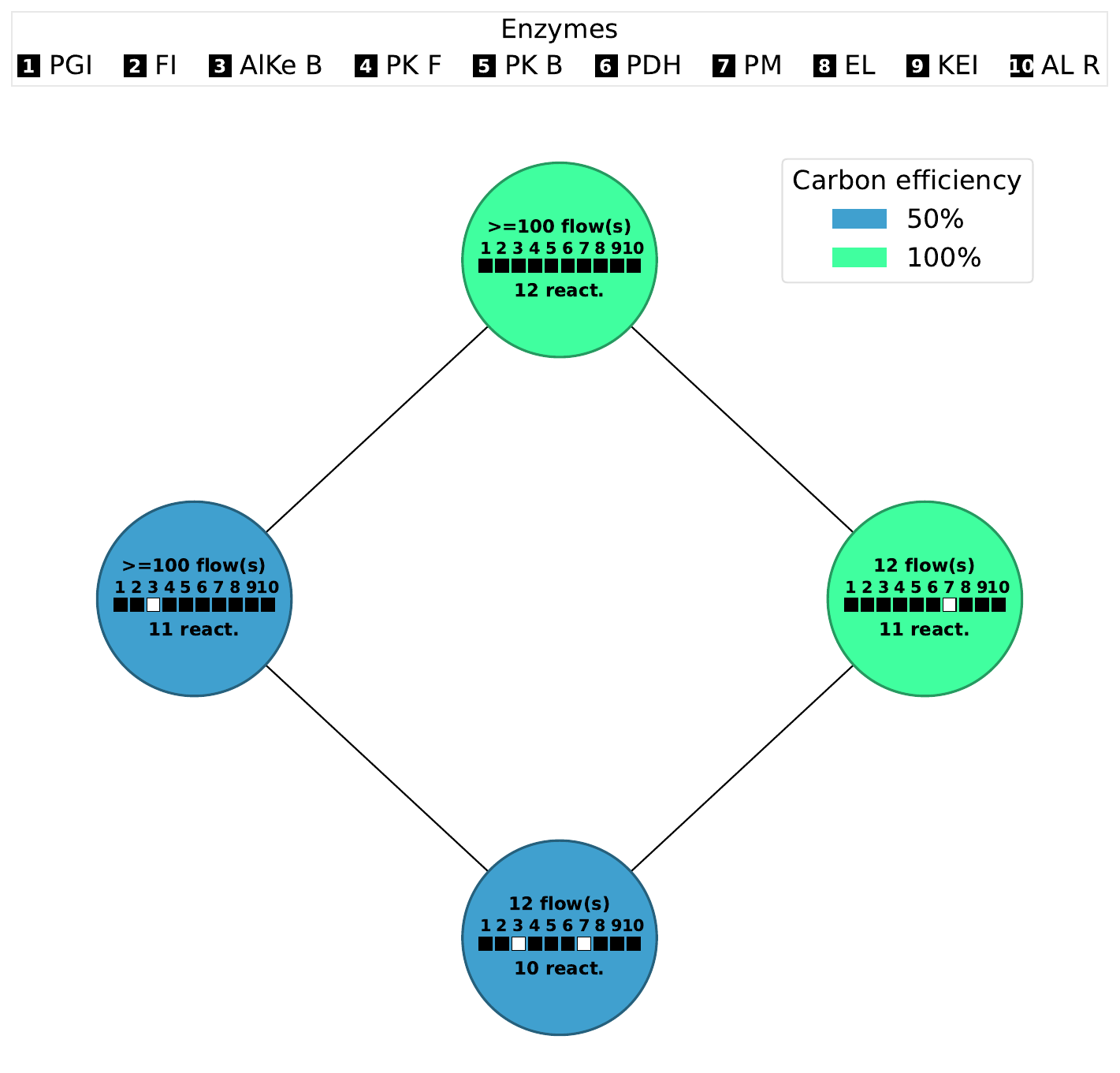}
    \caption{The Hasse diagram showing the results of applying the FGS method to Glycolysis. The Hasse diagram for the SGS method is identical. Note that we enumerated up to 100 pathways for each subset. Therefore, there most likely exists more than 100 solutions for the top node and the node to the right.}
    \label{fig:Glycolysis_wholespace}
    \end{figure}

    Fig.~\ref{fig:Glycolysis_wholespace} shows the Hasse diagram for the subsets that admit a pathway under each of the schemes. The Hasse diagrams for the FGS and SGS settings are identical. We observe that half of the nodes are $100\%$ carbon efficient. These are the subsets where AlKe B (enzyme number 3) is available, indicating that this enzyme is important for efficiency reasons, but not necessary for feasibility. 

    We observe that the enzyme sets where PM (enzyme number 7) is present produce a large amount of possible pathways. This is due to PM's high combinatorial potential. The intramolecular transfer of a phosphate group in a space with carbohydrates leads to a large number of possible reactions, since every free hydroxyl group is a potential acceptor. In our model, this leads to many phosphorylated sugars that can be used as an intermediate molecule that ends up in the final pathway.

    Additionally, we observe that one enzyme subset is just as efficient as the full set, namely the subset $\{\text{PGI, FI, AlKe B, PK F, PK B, PDH, EL, KEI, AL R}\}$ (enzyme numbers 1, 2, 3, 4, 5, 6, 8, 9, and 10). Notably, the only enzyme missing from this subset is PM. This follows the same logic as described before. In the way our modeling approach works, the correct configuration of phosphorylation can be present in the space already, therefore making the use of PM in the pathway unnecessary.
    What these alternative routes in modeling can represent is a margin for enzyme promiscuity. Many isoforms of enzymes across different organisms have different affinities for promiscuity and alternative products. These results give inspiration to explore the possibility of promiscuous reactions in the future for pathway design.


\subsection{General Findings}
Across the three systems, the feasible region of the powerset is consistently sparse, but the pattern of sparsity differs markedly between pathways. Non-oxPPP and especially Glycolysis admit only a small fraction of feasible subsets and tend to require relatively large enzyme combinations, whereas NOG retains feasibility across a broader range of subset sizes. At the same time, larger subsets are not systematically more carbon efficient: adding an enzyme may preserve, improve, or reduce the carbon efficiency because every enzyme in the subset is enforced to participate in the pathway. Taken together, this shows that pathway performance is governed less by subset size alone, and more by the specific functional combination of enzymes. The Hasse diagram structure therefore provides a natural way to identify dependencies, redundancies, and reduced high-performing pathway designs.

\subsection{A Local Perturbation View of the Hasse Diagram}

The analysis presented above characterizes each rule subset
independently by asking whether it can realize a desired pathway and, if so, how efficiently. The Hasse diagram contains additional information that is useful to understand how the subset metrics change as we move along the edges. Because these edges connect mechanisms that differ by
exactly one rule, we can analyze the local effect of introducing that rule.

We consider the NOG system, which is the case study that exhibits the largest Hasse diagram. The goal is then to characterize the effect of adding each rule, with respect to both feasibility and carbon efficiency.

For a fixed rule $r_i$, we can consider the edge from every subset $s$ not containing $r_i$ to $s \cup \{r_i\}$. Each transition edge therefore
corresponds to the introduction of exactly that one rule. Let the feasibility of a
subset be represented by a binary variable, where $0$ denotes an infeasible
subset and $1$ a feasible one. Every transition edge belongs to exactly one of four types:
$$
0 \rightarrow 0,\qquad
0 \rightarrow 1,\qquad
1 \rightarrow 0,\qquad
1 \rightarrow 1.
$$
For each rule $r_i$, we denote by
$$
n_{00},\qquad
n_{01},\qquad
n_{10},\qquad
n_{11}
$$
the number of transition edges of each type obtained when adding said rule. The first index denotes the feasibility before adding $r_i$, and the
second the feasibility afterward.


A first question is how frequently each rule appears among feasible subsets. Table~\ref{tab:nog-rule-prevalence} shows this quantity. As
expected, the PK rule occurs in every feasible subset, while other rules
appear in smaller fractions of them. 

\begin{table}[htbp]
    \centering
    \small
    \renewcommand{\arraystretch}{1.25}
    \begin{tabular}{@{}lcc@{}}
        \toprule
        \textbf{Rule} & {\textbf{\# feasible subsets}} & \textbf{Proportion of feasible subsets} \\
        \midrule
        AlKe B & 9  & 36\%  \\
        AlKe F & 21 & 84\%  \\
        TKL    & 14 & 56\%  \\
        TAL    & 14 & 56\%  \\
        AL     & 14 & 56\%  \\
        PHL    & 7  & 28\%  \\
        PK     & 25 & 100\% \\
        \bottomrule
    \end{tabular}
    \caption{Number of feasible NOG rule subsets containing each rule
    under FGS. A total of 25 feasible subsets were found.}
    \label{tab:nog-rule-prevalence}
\end{table}

\begin{figure}[H]
    \centering
    \includegraphics[width=.65\textwidth]{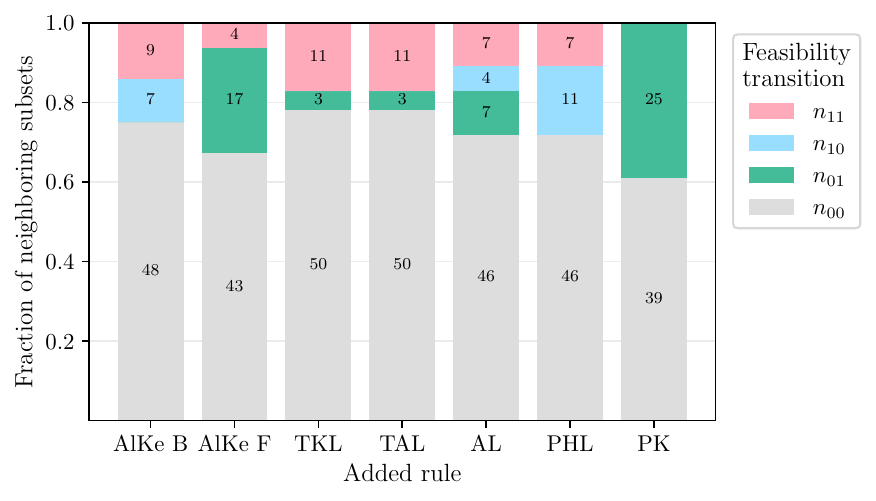}
    \caption{Feasibility transition plot. The colored segments indicate the number of cases in which adding the rule preserves
infeasibility ($n_{00}$), creates feasibility ($n_{01}$),
destroys feasibility ($n_{10}$), or preserves feasibility
($n_{11}$).}
    \label{fig:rule-transitions}
\end{figure}

Now, for each rule we examine the feasibility transitions, summarized in Figure~\ref{fig:rule-transitions}.
Every bar sums up to $2^6=64$, which is the number of subsets not containing rule $r_i$. The colored segments indicate the number of cases in which adding the rule preserves
infeasibility ($n_{00}$), creates feasibility ($n_{01}$),
destroys feasibility ($n_{10}$), or preserves feasibility
($n_{11}$).

This figure highlights the fact that different rules exhibit distinct structural roles. All rules mostly preserve infeasibility when added. However, among the remaining cases, there are some interesting behaviors worth emphasizing. PK and AlKe F predominantly enable new feasible subsets. Since PK is the only rule that can itself produce the target molecule from the input molecule, be it a low carbon efficiency, it is to be expected that the addition of the PK rule strictly creates feasibility. The AlKe F rule, by representing the reaction from an aldehyde to a ketone, can provide the prerequisites for other rules to apply, therefore enabling feasibility.

Rules like AlKe B, AL, and PHL can destroy feasibility when introduced, because the enlarged subset can no longer satisfy the constraint that all rules must be used at least once.
The remaining TKL and TAL rules only enable three new feasible subsets each, otherwise having no effect on feasibility. We note that the sum $n_{01} + n_{11}$ equals the number of feasible subsets that contain each rule, represented in Table~\ref{tab:nog-rule-prevalence}.

A natural next question is whether the addition of a rule changes the quality of the resulting pathway.
Carbon efficiency can only be compared between neighboring feasible subsets. We therefore restrict attention to the $1 \rightarrow 1$ transitions. Let $C(s)$ denote the carbon efficiency value for a given subset $s$. For each rule $r_i$ and each $1 \rightarrow 1$ transition, we compute the change in carbon efficiency
$$
\Delta C =
C(s \cup \{r_i\}) - C(s),
$$
and classify the observed changes according to their values.

\begin{figure}[H]
    \centering
    \includegraphics[width=.65\textwidth]{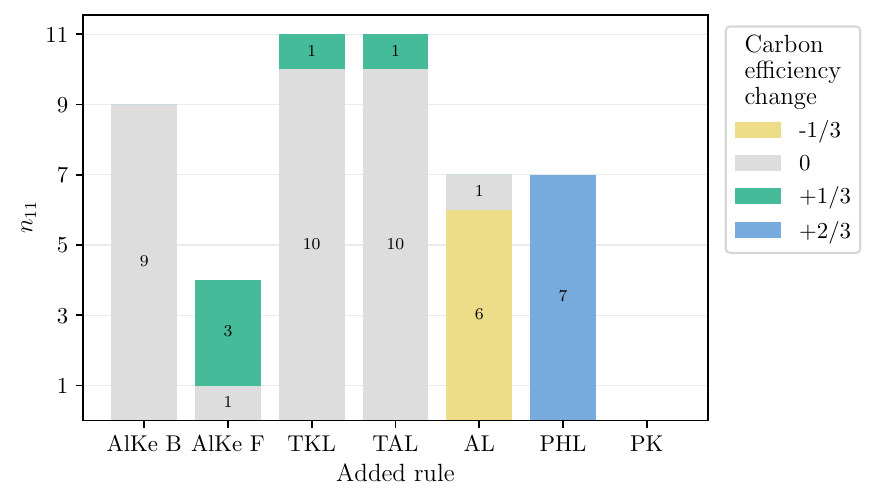}
    \caption{Carbon efficiency perturbation plot of the $1 \rightarrow 1$ transitions preserving feasibility (pink segments in Figure~\ref{fig:rule-transitions}). The
    height of each bar equals $n_{11}$, while the colored segments
    indicate the observed changes in carbon efficiency among the corresponding
    feasible neighboring subsets.}
    \label{fig:carbon-perturbations}
\end{figure}

Figure~\ref{fig:carbon-perturbations} summarizes these perturbations. Now the height of each bar equals $n_{11}$. The
colored segments indicate how many of these rule addition transitions improve, preserve, or
reduce carbon efficiency, and by how much.

Notably, every time rule PHL has been added to a feasible subset producing an enlarged also feasible subset, it has just improved the carbon efficiency values by $2/3$. 
This is especially interesting when comparing this behaviors with the feasibility transitions in figure \ref{fig:rule-transitions}, where the addition of PHL leads to either conservation of feasibility or, in the majority, to the destruction of feasibility. It seems that, when added to a feasible subset, the PHL rule dephosphorylates a bisphosphorylated molecule. When the carbon chain of that molecule is long enough, i.e. 7 carbons, the PK rule can apply twice to the resulting molecule, therefore increasing carbon efficiency by 2 AcP molecules.

In contrast, rule AL, representing an aldol addition reaction, has mainly decreased the carbon efficiency of the pathways when it has been added. This can be explained by the chain-elongation function of the rule, since the conversion of F6P to AcP is inherently a degradation pathway, with F6P consisting of a 6 carbon backbone and AcP of a 2 carbon backbone. The imposed constraint of applying every rule in the subset forces the use of the AL rule when present and can create a non-productive cycle in smaller subsets.

The rest of the rules have no outstanding effects in this metric. Finally, PK has no $1 \rightarrow 1$ transitions. 

\section{Conclusion and Outlook}
In this work, we presented a graph-theoretic framework for comparing all enzyme subsets of a biochemical system by combining chemical-space expansion, integer-hyperflow pathway search, and Hasse diagram organization of the feasible subsets. Applied to non-oxPPP, NOG, and Glycolysis, the framework reveals distinct structural signatures: non-oxPPP has few feasible subsets and a strong dependence on key rearrangement chemistry, NOG contains several reduced subsets that remain fully carbon efficient, and Glycolysis is markedly constrained, with feasibility concentrated near the full enzyme set. 

These results suggest that systematic subset analysis can serve as input for future pathway design work by highlighting minimal working sets, candidate shortcut pathways, and enzyme combinations. Hence, the methodology can act as a first approximation of which pathway candidates are worth investigating further in more detailed biochemical or experimental models, e.g. by including measures for thermodynamic feasibility and exploring enzyme availability.



\clearpage
\bibliographystyle{unsrt}
\bibliography{bibliography}

@article{flow,
    title = {Chemical Transformation Motifs --- Modelling Pathways as Integer Hyperflows},
    author = {Andersen, Jakob L. and Flamm, Christoph and Merkle, Daniel and Stadler, Peter F.},
    journal = {IEEE/ACM Transactions on Computational Biology and Bioinformatics},
    year = {2019}, 
    volume = {16}, 
    number = {2}, 
    pages = {510-523}, 
    doi = {10.1109/TCBB.2017.2781724},
    publisher = {IEEE},
    ISSN = {1545-5963}, 
    month = {March},
}

@InProceedings{mod_paper,
author="Andersen, Jakob L.
and Flamm, Christoph
and Merkle, Daniel
and Stadler, Peter F.",
editor="Echahed, Rachid
and Minas, Mark",
title="A Software Package for Chemically Inspired Graph Transformation",
booktitle="Graph Transformation",
year="2016",
publisher="Springer International Publishing",
address="Cham",
pages="73--88",
isbn="978-3-319-40530-8"
}

@InProceedings{50_shades_andersen,
author="Andersen, Jakob Lykke
and Flamm, Christoph
and Merkle, Daniel
and Stadler, Peter F.",
editor="Fages, Fran{\c{c}}ois
and Piazza, Carla",
title="50 Shades of Rule Composition",
booktitle="Formal Methods in Macro-Biology",
year="2014",
doi="10.1007/978-3-319-10398-3_9",
publisher="Springer International Publishing",
address="Cham",
pages="117--135",
isbn="978-3-319-10398-3"
}

@incollection{DBLP:conf/gg/CorradiniMREHL97,
  author       = {Andrea Corradini and
                  Ugo Montanari and
                  Francesca Rossi and
                  Hartmut Ehrig and
                  Reiko Heckel and
                  Michael L{\"{o}}we},
  editor       = {Grzegorz Rozenberg},
  title        = {Algebraic Approaches to Graph Transformation - Part {I:} Basic Concepts
                  and Double Pushout Approach},
  booktitle    = {Handbook of Graph Grammars and Computing by Graph Transformations,
                  Volume 1: Foundations},
  pages        = {163--246},
  publisher    = {World Scientific},
  year         = {1997},
  bibsource    = {dblp computer science bibliography, https://dblp.org}
}

@article{schwander_synthetic_2016,
	title = {A synthetic pathway for the fixation of carbon dioxide in vitro},
	volume = {354},
	url = {https://www.science.org/doi/10.1126/science.aah5237},
	doi = {10.1126/science.aah5237},
	number = {6314},
	urldate = {2024-03-13},
	journal = {Science},
	author = {Schwander, Thomas and Schada von Borzyskowski, Lennart and Burgener, Simon and Cortina, Niña Socorro and Erb, Tobias J.},
	month = nov,
	year = {2016},
	note = {Publisher: American Association for the Advancement of Science},
	pages = {900--904},
}

@article{mclean_2023_hopac,
author = {Richard P. McLean  and Thomas Schwander  and Christoph Diehl  and Niña Socorro Cortina  and Nicole Paczia  and Jan Zarzycki  and Tobias J. Erb },
title = {Exploring alternative pathways for the in vitro establishment of the {HOPAC} cycle for synthetic {CO}\textsubscript{2} fixation},
journal = {Science Advances},
volume = {9},
number = {24},
pages = {eadh4299},
year = {2023},
doi = {10.1126/sciadv.adh4299},
URL = {https://www.science.org/doi/abs/10.1126/sciadv.adh4299},
eprint = {https://www.science.org/doi/pdf/10.1126/sciadv.adh4299}}

@ARTICLE{Bogorad_2013-nog,
  title     = "Synthetic non-oxidative glycolysis enables complete carbon
               conservation",
  author    = "Bogorad, Igor W and Lin, Tzu-Shyang and Liao, James C",
  journal   = "Nature",
  publisher = "Springer Science and Business Media LLC",
  volume    =  502,
  number    =  7473,
  pages     = "693--697",
  doi       = "10.1038/nature12575",
  url       = "https://doi.org/10.1038/nature12575",
  month     =  oct,
  year      =  2013,
  language  = "en"
}

@article{hellgren_promiscuous_2020,
	title = {Promiscuous phosphoketolase and metabolic rewiring enables novel non-oxidative glycolysis in yeast for high-yield production of acetyl-{CoA} derived products},
	volume = {62},
	issn = {1096-7184},
	doi = {10.1016/j.ymben.2020.09.003},
	pages = {150--160},
	journal = {Metabolic Engineering},
	journaltitle = {Metabolic Engineering},
	shortjournal = {Metab Eng},
	author = {Hellgren, John and Godina, Alexei and Nielsen, Jens and Siewers, Verena},
	year={2020},
    date = {2020-11},
	pmid = {32911054},
}

@article{glycolysis_2026_gruening_review,
author = {Grüning, Nana-Maria and Agostini, Federica and Caldana, Camila and Hartl, Johannes and Heinemann, Matthias and Keller, Markus A. and Krüsemann, Jan Lukas and Lamperti, Costanza and Linster, Carole L. and Lindner, Steffen N. and Muenzner, Julia and Nielsen, Jens and Nikoloski, Zoran and Siebers, Bettina and Snoep, Jacky L. and Tenenboim, Hezi and Teusink, Bas and Williams, Spencer J. and Wamelink, Mirjam M. C. and Ralser, Markus},
title = {The return of metabolism: biochemistry and physiology of glycolysis},
journal = {Biological Reviews},
volume = {101},
number = {2},
pages = {751-803},
doi = {10.1111/brv.70104},
date = {2025-11-30},
url = {https://onlinelibrary.wiley.com/doi/abs/10.1111/brv.70104},
eprint = {https://onlinelibrary.wiley.com/doi/pdf/10.1111/brv.70104},
year = {2026}
}

@article{PPP_review_stincone_2015,
author = {Stincone, Anna and Prigione, Alessandro and Cramer, Thorsten and Wamelink, Mirjam M. C. and Campbell, Kate and Cheung, Eric and Olin-Sandoval, Viridiana and Grüning, Nana-Maria and Krüger, Antje and Tauqeer Alam, Mohammad and Keller, Markus A. and Breitenbach, Michael and Brindle, Kevin M. and Rabinowitz, Joshua D. and Ralser, Markus},
title = {The return of metabolism: biochemistry and physiology of the pentose phosphate pathway},
journal = {Biological Reviews},
volume = {90},
number = {3},
pages = {927-963},
doi = {10.1111/brv.12140},
url = {https://onlinelibrary.wiley.com/doi/abs/10.1111/brv.12140},
eprint = {https://onlinelibrary.wiley.com/doi/pdf/10.1111/brv.12140},
year = {2015}
}

@article{planet_compatible_chemindustry_2023,
author = {Fanran Meng  and Andreas Wagner  and Alexandre B. Kremer  and Daisuke Kanazawa  and Jane J. Leung  and Peter Goult  and Min Guan  and Sophie Herrmann  and Eveline Speelman  and Pim Sauter  and Shajeeshan Lingeswaran  and Martin M. Stuchtey  and Katja Hansen  and Eric Masanet  and André C. Serrenho  and Naoko Ishii  and Yasunori Kikuchi  and Jonathan M. Cullen },
title = {Planet-compatible pathways for transitioning the chemical industry},
journal = {Proceedings of the National Academy of Sciences},
volume = {120},
number = {8},
pages = {e2218294120},
year = {2023},
doi = {10.1073/pnas.2218294120},
URL = {https://www.pnas.org/doi/abs/10.1073/pnas.2218294120},
eprint = {https://www.pnas.org/doi/pdf/10.1073/pnas.2218294120}}

@article{rasor_toward_2021,
	title = {Toward sustainable, cell-free biomanufacturing},
	volume = {69},
	issn = {0958-1669},
	url = {https://www.sciencedirect.com/science/article/pii/S095816692030197X},
	doi = {10.1016/j.copbio.2020.12.012},
	series = {Chemical Biotechnology ● Pharmaceutical Biotechnology},
	pages = {136--144},
	journal = {Current Opinion in Biotechnology},
	journaltitle = {Current Opinion in Biotechnology},
	shortjournal = {Current Opinion in Biotechnology},
	author = {Rasor, Blake J and Vögeli, Bastian and Landwehr, Grant M and Bogart, Jonathan W and Karim, Ashty S and Jewett, Michael C},
	year = {2021},
	urldate = {2026-06-12},
	date = {2021-06-01}
}

@article{ko_tools_2020,
	title = {Tools and strategies of systems metabolic engineering for the development of microbial cell factories for chemical production},
	volume = {49},
	issn = {1460-4744},
	url = {https://pubs.rsc.org/en/content/articlelanding/2020/cs/d0cs00155d},
	doi = {10.1039/D0CS00155D},
	pages = {4615--4636},
	number = {14},
	journal = {Chemical Society Reviews},
	journaltitle = {Chemical Society Reviews},
	shortjournal = {Chem. Soc. Rev.},
	publisher = {The Royal Society of Chemistry},
	author = {Ko, Yoo-Sung and Kim, Je Woong and Lee, Jong An and Han, Taehee and Kim, Gi Bae and Park, Jeong Eum and Lee, Sang Yup},
	year = {2020},
	urldate = {2026-06-15},
	date = {2020-07-21},
	langid = {english},
}

@article{kim_metabolic_2023,
	title = {Metabolic engineering for sustainability and health},
	volume = {41},
	issn = {0167-7799, 1879-3096},
	url = {https://www.cell.com/trends/biotechnology/abstract/S0167-7799(22)00342-0},
	doi = {10.1016/j.tibtech.2022.12.014},
	pages = {425--451},
	number = {3},
	journal = {Trends in Biotechnology},
	journaltitle = {Trends in Biotechnology},
	shortjournal = {Trends in Biotechnology},
	publisher = {Elsevier},
	author = {Kim, Gi Bae and Choi, So Young and Cho, In Jin and Ahn, Da-Hee and Lee, Sang Yup},
	year = {2023},
	urldate = {2026-06-15},
	date = {2023-03-01},
}

@Misc{mod,
  Title                    = {{M\O D}},
  Author                   = {Jakob L. Andersen},
  howpublished             = {\url{http://mod.imada.sdu.dk}},
  Year                     = {2026},
}

@InCollection{Zeigarnik:00a,
  Title                    = {On Hypercycles and Hypercircuits in Hypergraphs},
  Author                   = {A. V. Zeigarnik},
  Booktitle                = {Discrete Mathematical Chemistry},
  Publisher                = {American Mathematical Society},
  Year                     = {2000},

  Address                  = {Providence, RI},
  Editor                   = {P. Hansen and P. W. Fowler and M. Zheng},
  Pages                    = {377--383},
  Series                   = {DIMACS series in discrete mathematics and theoretical computer science},
  Volume                   = {51},
}

@article{Muller:22,
author={Müller,Stefan and Flamm,Christoph and Stadler,Peter F.},
year={2022},
title={What makes a reaction network “chemical”?},
journal={Journal of cheminformatics},
volume={14},
number={1},
pages={63-63},
isbn={1758-2946},
language={English},
}

@article{Andersen:20,
    author = {Andersen, Jakob L. and Flamm, Christoph and Merkle, Daniel and Stadler, Peter F.},
    title = {Defining Autocatalysis in Chemical Reaction Networks},
    year = {2020},
    publisher = {nls - natural & life science publishers},
    address = {Switzerland},
    volume = {8},
    pages = {121--133},
    issn = {2571-7715},
    journal = {Journal of Systems Chemistry},
    url = {http://www.nls-publishers.com/shop/journal/journal+of+systems+chemistry+2020%2c+volume+8},
    note = {TR: \url{https://arxiv.org/abs/2107.03086}},
}

\clearpage
\appendix

\section{Full Enzyme Diagrams}

\begin{figure}[H]
    \centering
    \rotatebox{90}{%
        \includegraphics[width=\textheight]{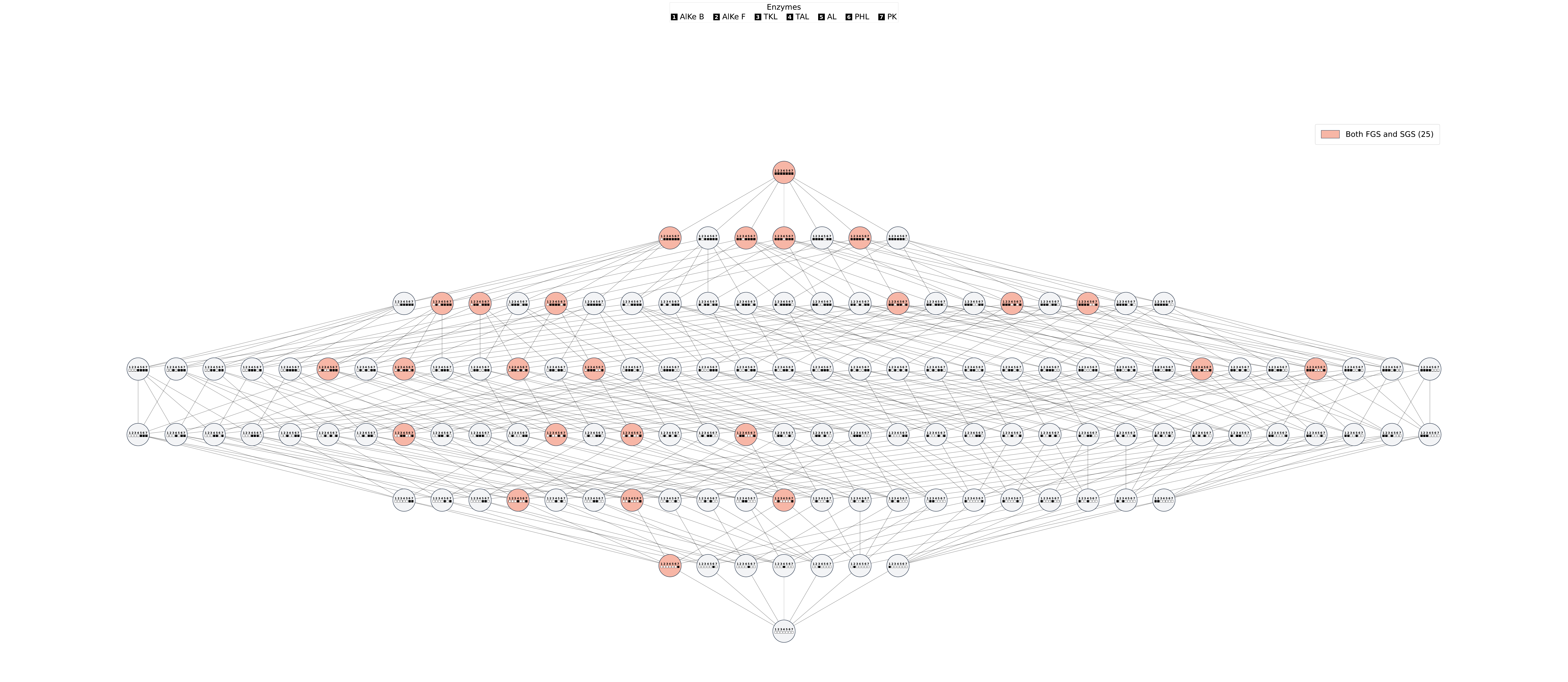}%
    }
    \caption{Nodes that admit a pathway while enforcing the use of all their enzymes for NOG.}
    \label{fig:NOG_feasible_region}
\end{figure}


\begin{figure}[H]
    \centering
    \rotatebox{90}{%
        \includegraphics[width=\textheight]{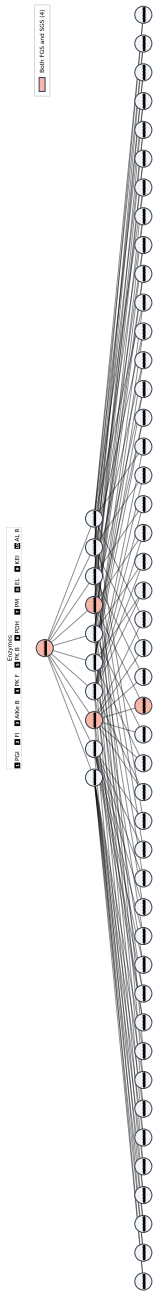}%
    }
    \caption{Nodes that admit a pathway while enforcing the use of all their enzymes for Glycolysis. We only visualize the levels where a pathway can be found.}
    \label{fig:Glycolysis_full_lattice}
\end{figure}

\clearpage
\section{Chemical Reaction Networks as Directed Hypergraphs}
A chemical reaction network (CRN) can be modeled by a directed hypergraph \cite{flow, Andersen:20, Muller:22, Zeigarnik:00a} where each molecule is represented by a vertex and each reaction is modeled by a directed hyperedge. For the formalisation of the concept we use \cite{flow} with slightly adapted notation. A directed hypergraph $\mathcal{H}=(V,E)$, consists of a vertex set $V$ and an edge set $E$. Each directed hyperedge $e \in E$ connects a pair of multisets of vertices, i.e. $e=(e^{\mathit{tail}}, e^{\mathit{head}})$ where $e^{\mathit{tail}}, e^{\mathit{head}} \subseteq V$.

    Additionally, the directed multi-hypergraph can be extended to allow for modeling of input and output reactions, called the \textit{extended hypergraph} \cite{flow} denoted by $\overline{\mathcal{H}}=(V,\overline{E}$ where $\overline{E}=E\cup E^{\mathit{in}} \cup E^{\mathit{out}}$ where
    \begin{align}
        E^{\mathit{in}} &= \{e^{\mathit{in}}_v = (\emptyset, \mset{v}) \mid v\in V\},\\
        E^{\mathit{out}} &= \{e^{\mathit{out}}_v=(\mset{v}, \emptyset) \mid v \in V\}.
    \end{align}

\section{Pathways as Integer Hyperflows}
    \begin{definition}{\cite[adapted]{flow}}
    An integer hyperflow on $\overline{\mathcal{H}}$ is a function $f:\overline{E}\rightarrow{\mathbb{N}_0}$ satisfying, for each $v\in V$ the conservation constraint
    \begin{align}
        \sum_{e\in\delta^{out}_{\overline{E}}(v)} m_v(e^{\mathit{tail}})f(e) - \sum_{e\in\delta^{in}_{\overline{E}}(v)} m_v(e^{\mathit{head}})f(e) = 0
    \end{align}
    \end{definition}
    That is, the sum of the flow leaving a vertex must be equal to the sum entering that same vertex. 

\section{Additional Implementation Details}

When expanding a CRN using MØD~\cite{mod_paper, mod} it is possible to limit the size of the generated molecules as to ensure that the expansion will end. To that end, we have chosen to limit the maximum number of carbons in each molecule. We have chosen to let this number reflect that of the "standard" pathway. That is for non-oxPPP and Glycolysis we use seven and six carbons, respectively, as the maximum number in each molecule. For NOG we base it on~\cite{flow} and set it to eight.

For the pathway finding, we enforce additional constraints for the reactions in the flow. For NOG and non-oxPPP we disallow reactions where there are no phosphates on a carbon chain as well as aldolase reactions where both carbon chains are shorter than three. Additionally for NOG, we disallow PKL reactions where both reactants have fewer than three carbons. For Glycolysis, we restrict the PM reaction to only apply intramolecularly, phosphorylation to only occur on terminal carbons, and reactions to only occur on phosphorylated carbon chains except for the initial phosphorylation of glucose. 


\begin{table}

\caption{Abbreviations of molecule names defined in the chemical space.}
\label{tab:app_mol_names}
\centering
\begin{tabular}{lp{9cm}}
\toprule
     Abbreviation & Name \\ \midrule
    Ru5P & Ribulose-5-phosphate \\
    F6P & Fructose-6-phosphate \\
    Pi & Phosphate\\
    AcP & Acetyl-CoA \\
    G3P & Glyceraldehyde-3-phosphate \\
    DHAP & Dihydroxyacetone phosphate \\ 
    E4P & Erythrose 4-phosphate \\ 
    R5P & Ribose-5-phosphate \\
    S7P & Sedoheptulose-7-phosphate \\
    FBP & Fructose-1,6-phosphate \\
    G6P & Glucose-6-phosphate \\
    C8P & 8-carbon-phosphorylated-sugar \\
\bottomrule
\end{tabular}
\end{table}

\section{Rule Definitions}

\subsection{Non-Oxidative Pentose Phosphate Pathway Rules}

\begin{figure}[htpb]

    \centering
    \powersetRuleBlock{AlKe B}
        {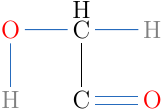}
        {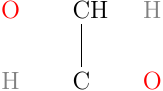}
        {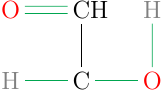}
        {%
    \par\smallskip
    \begin{minipage}{0.95\linewidth}
        {\footnotesize\itshape Constraint:}
        \vspace{-0.75\baselineskip}
        {\scriptsize\input{rules/002_r_0_constraints.tex}}
    \end{minipage}
}
    \powersetRuleBlock{AlKe F}
        {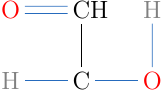}
        {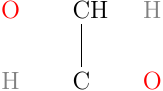}
        {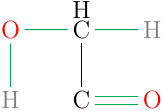}
        {%
    \par\smallskip
    \begin{minipage}{0.95\linewidth}
        {\footnotesize\itshape Constraint:}
        \vspace{-0.75\baselineskip}
        {\scriptsize\input{rules/005_r_1_constraints.tex}}
    \end{minipage}
}
        \end{figure}
\begin{figure}[htpb]

    \centering
    \powersetRuleBlock{TKL}
        {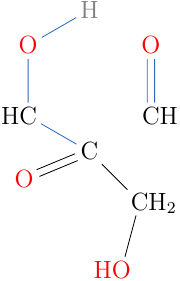}
        {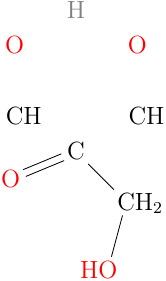}
        {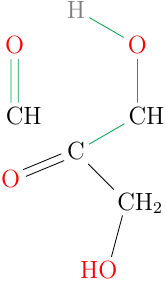}
        {}

    \powersetRuleBlock{TAL}
        {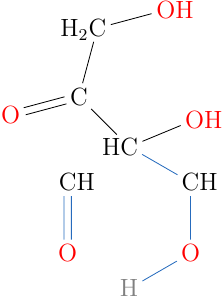}
        {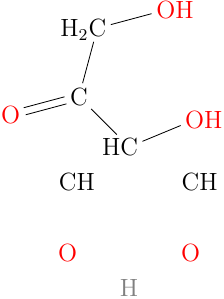}
        {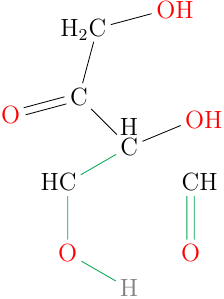}
        {}
    \powersetRuleBlock{AL}
        {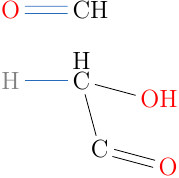}
        {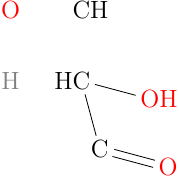}
        {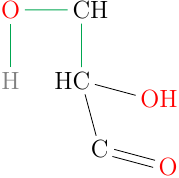}
        {}
    \powersetRuleBlock{PHL}
        {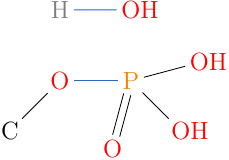}
        {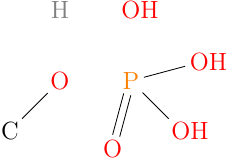}
        {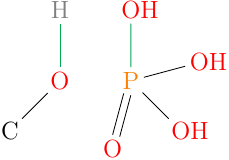}
        {}
\end{figure}

\FloatBarrier
\subsection{Non-Oxidative Glycolysis Rules}

The non-oxidative glycolysis case study extends the non-oxidative pentose
phosphate pathway rule set with the phosphoketolase rule shown below.

\begin{figure}[htpb]

    \centering
    \powersetRuleBlock{PK}
        {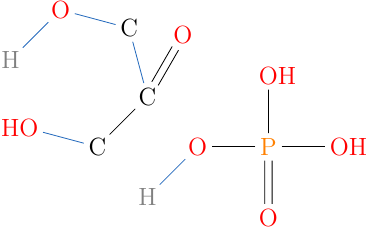}
        {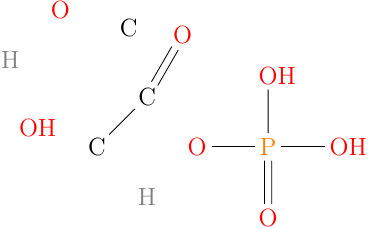}
        {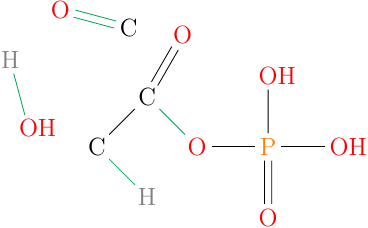}
        {}
\end{figure}

\FloatBarrier
\subsection{Glycolysis Rules}

\begin{figure}[htpb]

    \centering
    \powersetRuleBlock{PGI}
        {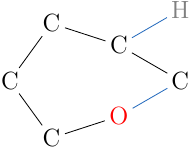}
        {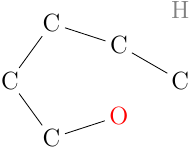}
        {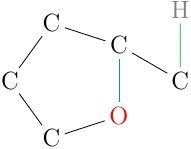}
        {}
    \powersetRuleBlock{FI}
        {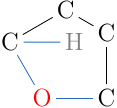}
        {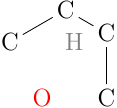}
        {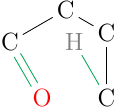}
        {}
    \powersetRuleBlock{AlKe B}
        {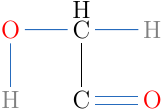}
        {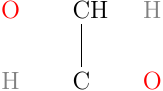}
        {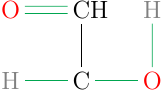}
        {%
    \par\smallskip
    \begin{minipage}{0.95\linewidth}
        {\footnotesize\itshape Constraint:}
        \vspace{-0.75\baselineskip}
        {\scriptsize\input{rules/047_r_15_constraints.tex}}
    \end{minipage}
}
\end{figure}

\begin{figure}[htpb]

    \centering
    \powersetRuleBlock{PK F}
        {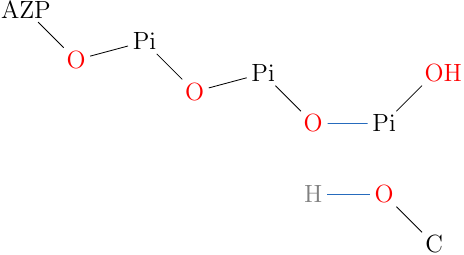}
        {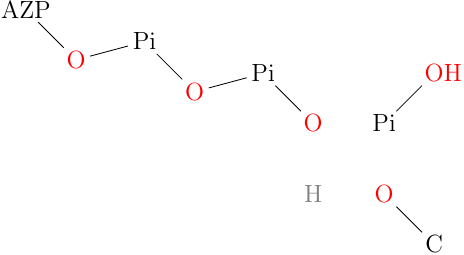}
        {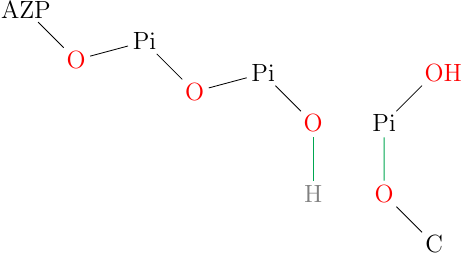}
        {%
    \par\smallskip
    \begin{minipage}{0.95\linewidth}
        {\footnotesize\itshape Constraint:}
        \vspace{-0.75\baselineskip}
        {\scriptsize\input{rules/050_r_16_constraints.tex}}
    \end{minipage}
}

    \powersetRuleBlock{PK B}
        {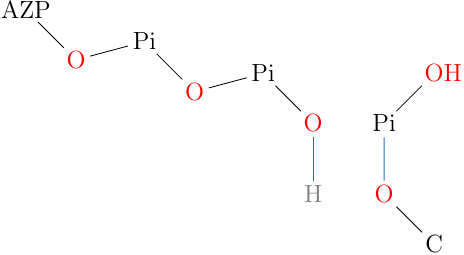}
        {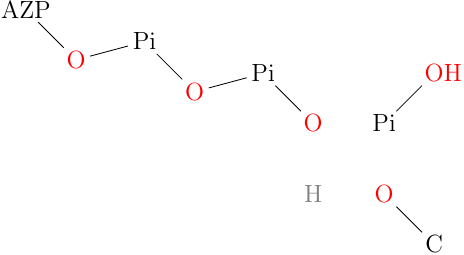}
        {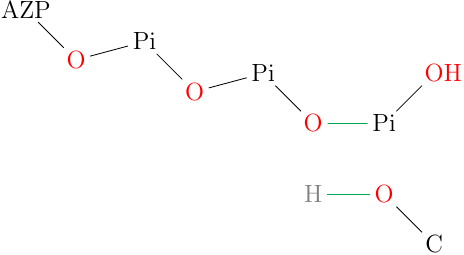}
        {}
    \powersetRuleBlock{PDH}
        {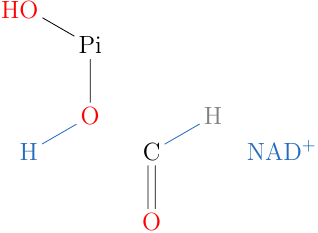}
        {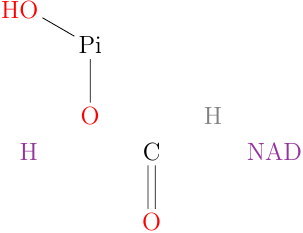}
        {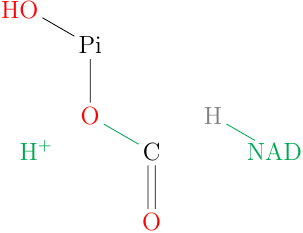}
        {}

    \powersetRuleBlock{PM}
        {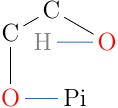}
        {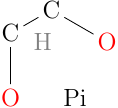}
        {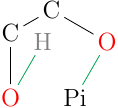}
        {}
    \powersetRuleBlock{EL}
        {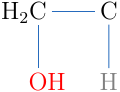}
        {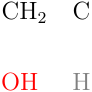}
        {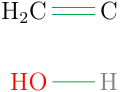}
        {}
    \powersetRuleBlock{KEI}
        {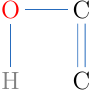}
        {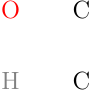}
        {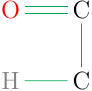}
        {}
    \powersetRuleBlock{AL R}
        {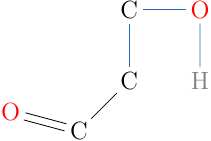}
        {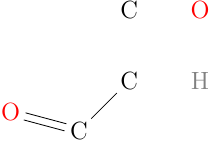}
        {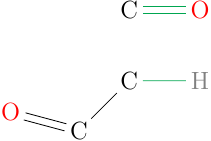}
        {}
\end{figure}

\end{document}